\documentclass{ar-1col}
\usepackage[numbers]{natbib}
\usepackage{graphicx}
\usepackage{url}
\usepackage{multirow}
\jname{Annu. Rev. Nucl. Part. Sci.}
\jvol{70}
\jyear{2020}
\doi{10.1146/annurev-nucl-020420-093547}

\newcommand{\pt}{\ensuremath{p_{\mathrm{T}}}}
\newcommand{\ifb}{\ensuremath{{\mathrm{fb}^{-1}}}}
\newcommand{\ttbar}{\ensuremath{t \bar{t}}}

\begin{document}

% Page header
\markboth{A. Ryd \& L. Skinnari}{Tracking Triggers for the HL-LHC}

% Title
\title{Tracking Triggers for the HL-LHC}

%Authors, affiliations address.
\author{Anders Ryd$^1$ and Louise Skinnari$^2$
\affil{$^1$Department of Physics, Cornell University, Ithaca, NY 14853, USA; \\ email: Anders.Ryd@cornell.edu}
\affil{$^2$Department of Physics, Northeastern University, Boston, MA 02115, USA; \\ email: l.skinnari@northeastern.edu}}

%Abstract
\begin{abstract}

Hardware-based track reconstruction in the CMS and ATLAS trigger systems for the High-Luminosity LHC upgrade will provide unique capabilities. 
An overview is presented of earlier track trigger systems at hadron colliders, in particular for
the Tevatron CDF and D\O\ experiments. 
We discuss the plans of the CMS and ATLAS experiments to implement hardware-based track reconstruction for the High-Luminosity LHC. 
Particular focus is placed on the track trigger capability of the upgraded CMS experiment. We discuss the challenges and opportunities of this novel handle, and review the alternatives that were considered for its implementation as well as discuss the expected performance. 
The planned track trigger systems for CMS and ATLAS have different goals, and we compare and contrast the two approaches. 

\end{abstract}

%Keywords, etc.
\maketitle

\tableofcontents

%%%%%%%%%%%%%%%%%%%%%%%%%%%%%%%%%%%%%%%%%%%%%%%%%%%
\section{INTRODUCTION}
\label{sec:introduction}

A critical component of any high-energy particle physics collider experiment is deciding which collision events to read out and save for future analysis, and which ones to discard. At the Large Hadron Collider (LHC)~\cite{Bruning:782076,Evans_2008} at CERN, proton-proton ($pp$) collisions occur every 25~ns, corresponding to a beam crossing frequency of 40~MHz. 
For the ATLAS~\cite{atlas} and CMS~\cite{cms} experiments, a typical collision event had a size of $\sim$1~MB during LHC Run~1--3. This corresponds to a data rate of roughly 40~TB per second, a rate much too large to both read out from the detectors and to store offline. 
Instead, only a small fraction of $pp$ collisions is kept, made possible by the comparatively low cross sections for the physics processes of interest compared to the total LHC cross section.  
The system utilized to decide which collision events to retain is referred to as the {\it trigger system}. It typically consists of an initial hardware-based level implemented in custom electronics boards, as well as a high-level trigger implemented in software. 

The High-Luminosity LHC (HL-LHC) upgrade~\cite{hllhc, hllhc_new}, planned for installation in 2025--2027, will significantly increase the instantaneous luminosity. As a consequence, a large increase in the number of simultaneous $pp$ interactions within the same bunch crossing, known as {\it pileup}, is also expected. This is a challenge particularly for the hardware-based trigger; novel handles are required for ATLAS and CMS to mitigate these effects. 

The identification of charged-particle trajectories (tracking) using silicon-based detectors is central to the LHC experiments. However, thus far, this has been restricted to software-based algorithms implemented in commercial CPUs in the high-level trigger or in the offline event reconstruction. In this review, we discuss the developments of hardware-based tracking (track triggering), specifically utilizing inputs from silicon tracking detectors. 
The focus of the review is the development that is underway for the CMS experiment for the HL-LHC operation to incorporate full-detector track reconstruction in the initial trigger level at an input rate of 40~MHz. 
We discuss the motivation and physics potential for utilizing hardware-based tracking, the associated challenges at the HL-LHC, and review prior developments in this area. Details are provided about the CMS track trigger system, including the design of the new outer tracker for HL-LHC, as well as the different track finding approaches that have been studied, along with their implementation. We discuss the foreseen HL-LHC track-finding system that will be based on field-programmable gate arrays (FPGAs), the hardware platforms to be used, as well as the expected performance. The plans for the ATLAS experiment to incorporate hardware-based tracking for HL-LHC are also reviewed. We discuss the foreseen new tracking detector, two alternative trigger architectures, and the approach identified for performing the track reconstruction. Finally, a summary and conclusion is given.

%%%%%%%%%%%%%%%%%%%%%%%%%%%%%%%%%%%%%%%%%%%%%%%%%%%
\subsection{Challenges and Requirements at the HL-LHC} 

The HL-LHC era offers exciting physics possibilities albeit with a substantially increased instantaneous luminosity, which is experimentally challenging. The detector systems will therefore undergo significant upgrades in order to maximally take advantage of the physics potential of the HL-LHC data sets. The main challenge of the HL-LHC operation for the ATLAS and CMS experiments is the large increase in the number of simultaneous $pp$ interactions. At the LHC design luminosity, an average of 25 $pp$ interactions occur in each bunch crossing, whereas at the HL-LHC, an average of 200 simultaneous interactions are expected. 

The large increase in pileup results in many more particles produced in each bunch collision, and consequently, an ambient increased energy in the calorimeter measurements, an increased number of low-momentum muons that could be misidentified as high-momentum ones, and so forth. Improving the experimental handles to identify particles in the trigger is thus required.
A second experimental challenge of the HL-LHC operation is high radiation levels that, in particular, the innermost detector systems must accommodate. Radiation tolerance is thus a key requirement in the design of the silicon tracker.
Consequently, the ATLAS and CMS experiments will improve the trigger systems, entirely replace the inner silicon-based tracking detectors, and upgrade large components of the readout systems associated with the muon and calorimeter systems. 

%%%%%%%%%%%%%%%%%%%%%%%%%%%%%%%%%%%%%%%%%%%%%%%%%%%
\subsection{Motivation for Hardware-Based Tracking} 
\label{sect:tracktriggermotivation}

The motivation for identifying charged-particle trajectories in a hardware-based trigger system is two-fold. First, it is a powerful handle to improve the identification of different types of particles and, consequently, to enable the physics goals of the experiments. Second, the inclusion of charged particle trajectories is a novel experimental handle in the trigger that can facilitate entirely new analyses in currently unexplored corners of phase space. 

The HL-LHC accelerator upgrade will increase the instantaneous luminosity for $pp$ collisions to roughly $7.5\times10^{34}$~cm$^{-2}$s$^{-1}$, about four times larger than the maximum instantaneous luminosity at the LHC. The target during the HL-LHC operation is to collect data sets of 3,000~\ifb\ for each of the ATLAS and CMS experiments, an order of magnitude larger than the full LHC data sets. The large data samples will enable detailed studies of rare Standard Model (SM) processes, precise Higgs boson ($H$) measurements, and extend sensitivity in searches for particles and interactions beyond the SM~\cite{ATLAS_LOI, ATLAS_SD, tp, CMS_SD, yellowreport}. 
Extensive studies of the $H$ boson production modes, properties, and interactions is a key goal. Rare decays, e.g. to $\mu^+\mu^-$ and $Z\gamma$, are expected to be conclusively observed. The $H$ boson self-coupling, which probes the Higgs field potential, will be studied through measuring the extremely rare $HH$ production process. Other important tests of electroweak symmetry breaking include measurements of electroweak multiboson interactions and studies of quartic boson couplings, all low cross section processes that are currently not well constrained. In the flavor sector, the increased integrated luminosity will particularly benefit measurements of rare $b$ and $c$ hadron decays, an indirect probe of beyond-SM physics, and can significantly improve searches for e.g. top quark flavor-changing neutral currents. Direct new physics searches, e.g. for signatures of supersymmetry, dark matter candidates, or new heavy gauge bosons, will be probed to higher mass scales. 

The above studies rely on the capability of the experiments to identify, at the trigger level, physics objects at the electroweak scale with high efficiency. Momentum thresholds must be maintained sufficiently low to capture leptons and hadronic $\tau$ decays from $W/Z/H$ boson decays and low transverse momentum (\pt) jets associated with $b$ quarks. The searches for new physics, e.g. supersymmetry, require the identification of multiple low-$\pt$ objects, searches for dark matter particles typically relies on the triggering capability of missing transverse momentum, and so forth.

The ATLAS and CMS trigger systems for LHC Run 1--3 read out merely about 0.25\% of the LHC collisions, or 100~kHz, from the initial hardware-based (Level-1, or L1) trigger. Following the software-based trigger, this data rate is reduced even further to about 1~kHz~\cite{atlas-trigger,cms-trigger}.
These L1 trigger systems utilize only information from the calorimeter and muon systems. 
To maintain the triggering capability at the HL-LHC while keeping the trigger rate at a manageable level, the trigger systems must be upgraded and improved. The inclusion of charged-particle tracking is a critical component in this. 
The precise measurement of a trajectory's transverse momentum improves the identification of muons, reducing the rate of low-$\pt$ muons that are misidentified as a high-$\pt$ ones; an example of this is illustrated for the CMS experiment in Fig.~\ref{fig:muonTP}. Track information also improves electron identification by matching tracks to calorimeter clusters, the reconstruction of hadronic $\tau$ decays, 
and allows a selection based on tracks of charged ($e, \mu, \tau$) and neutral ($\gamma$) particles that are isolated from other activity in the detector to reduce event rates.
Beyond improving the identification of individual trigger objects, tracking can be used in defining event-level quantities, e.g. the primary collision vertex to measure global quantities (e.g. the missing transverse momentum), and in correlating multiple objects when defining trigger signatures. 

\begin{figure}[bth]
\begin{center}
\begin{minipage}[t]{0.48\textwidth}
\includegraphics[width=\textwidth]{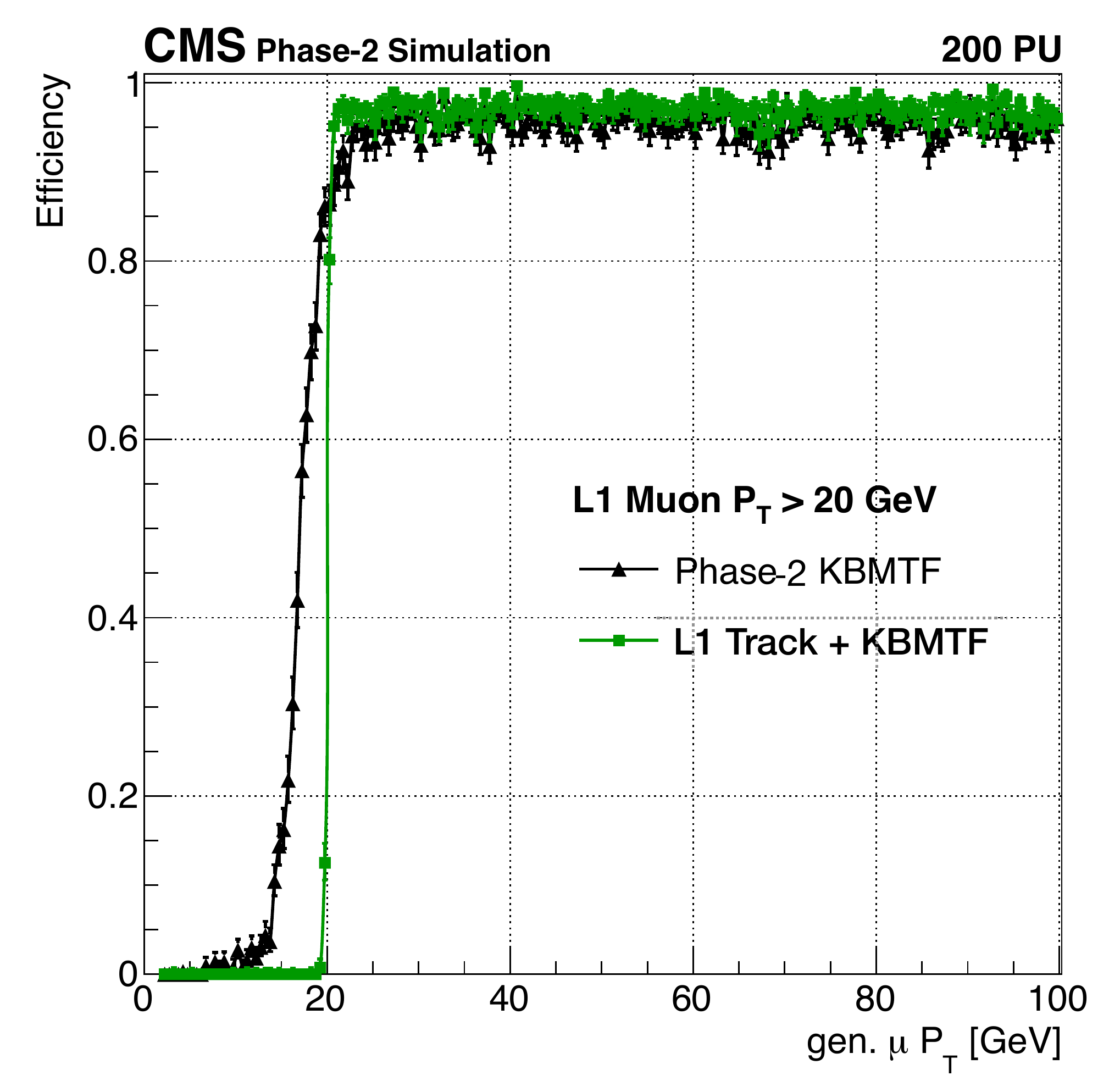}
\end{minipage}
\begin{minipage}[t]{0.51\textwidth}
\includegraphics[width=\textwidth]{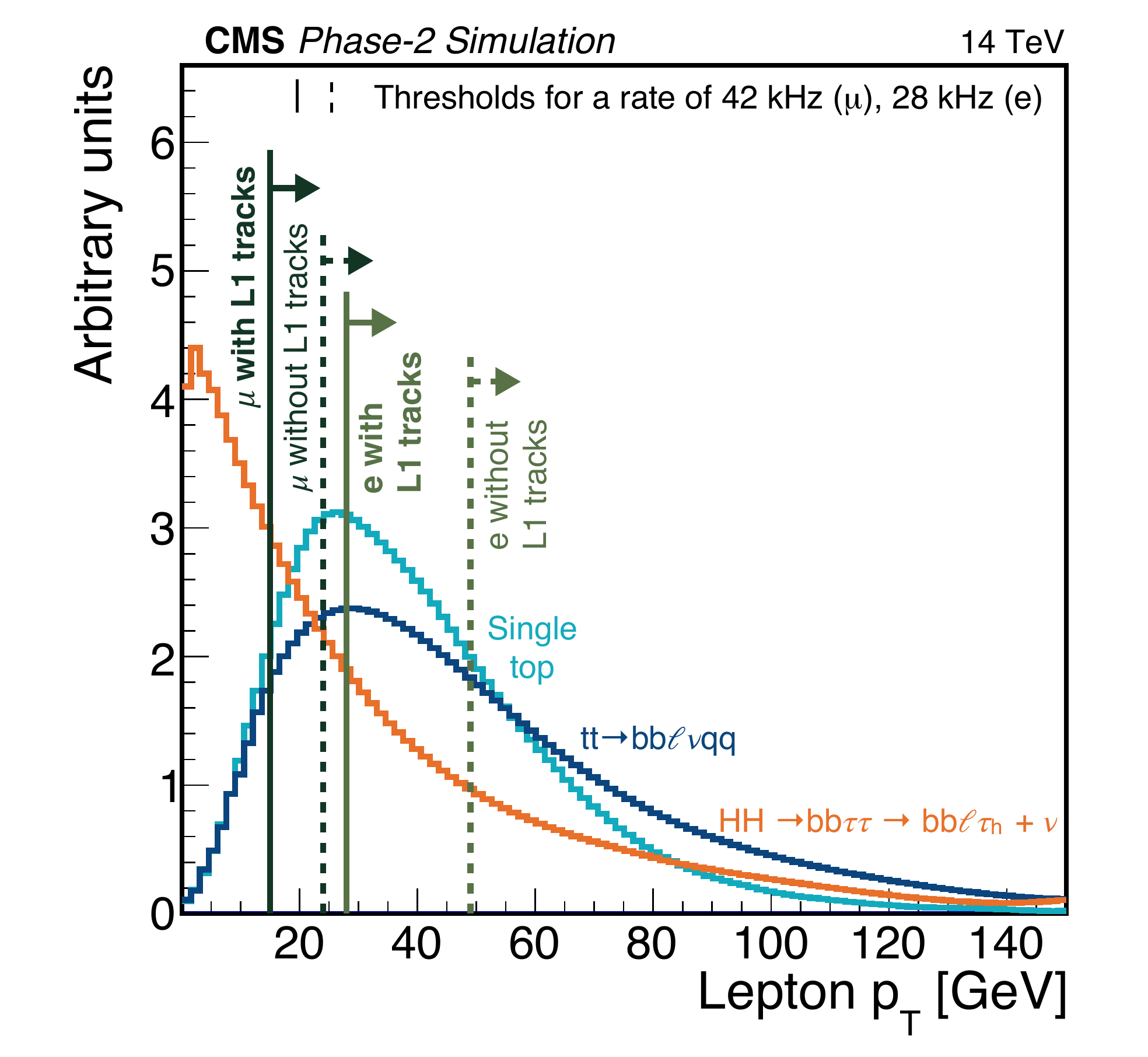}
\end{minipage}
\caption{Example of the power of including L1 tracking in CMS for the HL-LHC. The left figure illustrates the improved muon momentum resolution when including tracks (Track + KBMTF) for the barrel muon track finder (Phase-2 KBMTF). The right figure shows simulated electron and muon $\pt$ distributions from $HH$, single top quark, and semileptonic $t\bar{t}$ decays. The solid (dashed) vertical lines correspond to the trigger thresholds with (without) using L1 tracks for a given trigger rate, assuming 200 pileup interactions. The addition of tracking information reduces the trigger rate through the resulting improved object identification. From Ref.~\cite{CMS_P2_L1TDR}.}
\label{fig:muonTP}
\end{center}
\end{figure}

The inclusion of track reconstruction in the hardware-based trigger can also enable entirely new studies, probing physics processes that could previously not be identified in the trigger, e.g. due to overwhelming background processes when relying only on information from the calorimeter and muon system. Such signatures include e.g. displaced trajectories from hypothetical exotic long-lived particles and $B$ physics processes with low-$\pt$ final-state particles~\cite{TrackerTDR}. The full physics capability of incorporating tracking in the L1 trigger is yet to be explored.

The LHCb experiment, specialized for studying charm and beauty physics, is being upgraded during the LHC Long Shutdown 2 in 2019--2021~\cite{LHCbCollaboration:2014vzo}. Since the fraction of $pp$ collisions containing charm or beauty quarks is very high, and the instantaneous luminosity for LHCb is lower than for ATLAS and CMS, these upgrades use a different strategy for selecting the interesting collisions. LHCb is implementing a trigger-less system where all collisions at an input rate of about 30~MHz are read out from the detector, and the events are processed in the event filter farm, which is a fully software-based trigger.

%%%%%%%%%%%%%%%%%%%%%%%%%%%%%%%%%%%%%%%%%%%%%%%%%%%
\subsection{Early Developments}
\label{sect:earlydevelopments}

Track triggers for various applications have been used at many particle physics experiments.
Some examples include track identification in $e^+e^-$ experiments, e.g. BABAR and Belle~\cite{Bailey:2004rd,Iwasaki:2011za}, and
identification of long-lived particles, e.g. at HERA-B~\cite{Gerndt:2000dr}.
The early applications of track triggers that most closely relates to the challenges at the HL-LHC were used to perform track and vertex identification at the CDF and D\O\ \cite{cdf_svt, Anderson:2003pe} hadron collider experiments at the Tevatron proton-antiproton ($p\bar{p}$) collider at Fermilab. 
Developments for the ATLAS Fast TracKer (FTK) system~\cite{FTK} also provides important 
guidance toward the upgrades for the HL-LHC. Here, a brief review of the
earlier track trigger developments at hadron colliders is provided.

The CDF and D\O\ experiments were upgraded for Run 2 of the
Tevatron to provide additional experimental capabilities. In particular, this included new or
expanded capabilities for tracking and track triggering. Both experiments provided
tracking in their outer tracking detectors at the full bunch collision rate to 
provide improved triggering of muons and other particles. In addition, tracks in the 
outer trackers were combined with more precise hits from the silicon tracking
detectors to provide precise impact parameter measurements for the identification
of long-lived particles from e.g. $b$ hadrons.
The CDF and D\O\ track triggers operated in an environment with bunch 
crossings every 396~ns (2.5~MHz) and an average pileup of about 4 to 5. 
In comparison, at the 
HL-LHC bunch collisions will take place every 25~ns (40~MHz) with an average pileup 
of about 200.

The CDF II Silicon Vertex Tracker (SVT)~\cite{cdf_svt} made use of measurements from
the Silicon Vertex Tracker (SVXII) and the Central Outer Tracker (COT) to
reconstruct precise 2-D trajectories of charged particles. 
The CDF track trigger used two stages, the first was 
the eXtremely Fast Tracker (XFT)~\cite{cdf_xft,cdf_xft2} that used hits from the COT to
reconstruct charged particle trajectories. The XFT
first found track segments in four adjacent cells consistent with 
$\pt>1.5$~GeV. The segments were linked together to form complete 
tracks. The SVT in the second stage was implemented using Associative Memories (AM)~\cite{DellOrso:1990tzl} where the
input was hits from the SVXII detector and the XFT tracks. The AM-based
pattern recognition was implemented using 64 sectors in $\phi$. Each sector used
two AM boards that had 128 AM chips, each with 128 patterns for a total
of 32k patterns for a sector.  
The patterns
found in the AM were forwarded to the track fitting stage, implemented
in FPGAs. The track fit was performed using a linearized $\chi^2$ fit.

The D\O\ experiment implemented track reconstruction in their L1 Central Track Trigger (L1CTT)~\cite{Anderson:2003pe}
using hits from the Central Fiber Tracker (CFT)~\cite{d0_cft}. The 
track finding was implemented using FPGAs in 4.5$^{\circ}$ sectors. Each sector
used four FPGAs for track finding in different $\pt$ ranges ($>10$ GeV,
5 to 10 GeV, 3 to 5 GeV, and 1.5 to 3 GeV). A track was required to have
hits in all eight axial CFT layers. The pattern recognition was implemented
using combinatorial logic in the FPGA. The number of track equations
implemented in each FPGA varied from 3,000 for the highest $\pt$ range
to 10,000 for the lowest $\pt$ range. The tracks found in groups of ten
track finding sectors were collected in an octant board, which in turn found the
sectors with the highest occupancy and identified isolated tracks that 
were used for the L1 trigger decision with a latency of 2.5 $\mu$s with
respect to the beam collision time.
The L1 tracks were also used as input to the L2 track finding. The maximum
input rate for the L2 trigger was 10~kHz. The L2 processor boards were based on
1~GHz Pentium processors. The Level-2 Silicon Tracking Trigger (L2STT) used tracks found in the L1CTT and
added precise Silicon Microstrip Tracker hits. This allowed rejecting misreconstructed ("fake") L1CTT
tracks, improving the $\pt$ resolution, and most importantly it allowed identifying
tracks from long-lived particles, in particular $b$ hadrons~\cite{Adams:2007rr}, using 
the precise transverse impact parameter.

The ATLAS Collaboration developed a track trigger, FTK, for the Phase-1
upgrades. The FTK upgrade implemented global track 
reconstruction for events selected by the L1 trigger at a 
maximim rate of 100~kHz.. 
The design goal of the FTK was to reconstruct tracks with $\pt$ 
greater than 1~GeV.
The FTK used hits from the semiconductor tracker (SCT), the 
pixel detector, and the insertable B-layer (IBL). 
The hits were organized into 12 logical 
detector layers, eight in the SCT and four in the pixel+IBL. 
The FTK was implemented
using AMs for the pattern recognition,
using 8 of the 12 layers. In a second stage, the hits found
in the pattern recognition step
were used for the final track fit, including also hits matched to the track in the four
layers not used in the pattern recognition. The full ATLAS FTK system was
not installed, but a slice of the system was operated during the
Run 2 of the LHC~\cite{Sottocornola:2018sqx}.

%%%%%%%%%%%%%%%%%%%%%%%%%%%%%%%%%%%%%%%%%%%%%%%%%%%
\section{CMS TRACK TRIGGER} 
\label{sec:tracktrigger}

A key goal of the CMS detector upgrades for the HL-LHC operation is to 
implement track finding at the L1 trigger level 
to keep thresholds sufficiently low to maintain high efficiency for electroweak physics.
The goal of the CMS track trigger is to
reconstruct trajectories of charged particles with $\pt>2$~GeV for all $pp$ interactions. These 
tracks will be available in the L1 trigger and will provide much
improved identification of objects such as muons, electrons, taus, and
hadronic jets. To accomplish this goal, CMS has designed 
a silicon-based outer tracker~\cite{TrackerTDR} that is uniquely capable of producing trigger 
primitives, {\it stubs}, that are used to reconstruct the L1 tracks at 40~MHz.

%%%%%%%%%%%%%%%%%%%%%%%%%%%%%%%%%%%%%%%%%%%%%%%%%%%
\subsection{Outer Tracker for HL-LHC} 
\label{sect:tracker}

The CMS tracking detector will be replaced for the HL-LHC operation as the original
detector would not be able to handle the expected data rate or radiation dose. 
The complete replacement of the tracking detector provides a unique
opportunity to introduce new designs and capabilities. CMS has
used this opportunity to design a tracker capable of providing
$\pt$ discrimination at the detector module level before readout.
The CMS tracker for HL-LHC consists of an inner tracker (IT) with pixel sensors, which is not utilized in the L1 trigger, and an outer tracker (OT) with dedicated {\it $\pt$ modules}.
The ability to provide $\pt$ discrimination results in a sufficient
reduction of the data rate to allow readout of hits above a threshold
of about 2~GeV for use in the L1 trigger. This unique detector design is essential in enabling full-detector track reconstruction at the full 40~MHz bunch
crossing rate.

A charged particle produced at the interaction point follows a
trajectory in the transverse plane in a uniform magnetic field along the beam-axis
given by 
\begin{equation}
\phi=\phi_0+\arcsin\left(\frac{qB}{2p_{\rm T}}\cdot r\right)  
\end{equation} 
where $\phi$ is the azimuthal position of the trajectory at radius $r$, 
$\phi_0$ is the particle azimuthal direction at the origin, $B$ is the magnetic field
strength, $q$ is the particle charge, and $\pt$ is the transverse 
momentum. For two hits with a radial separation of $\Delta r$, this corresponds to
a $\phi$ separation
\begin{equation}
\Delta\phi=\Delta r\frac{qB}{2p_{\rm T}}  
\end{equation} 
where the approximation $\arcsin x=x$ has been used. The 
separation along the azimuthal direction $x$ is given by
\begin{equation}
\Delta x=r\Delta r\frac{qB}{2p_{\rm T}}.  
\end{equation}
The separation of hits is proportional to $r$, $\Delta
r$, $B$, and inversely proportional to $\pt$. 

The outer tracker design uses detector modules that consist of 
two silicon sensors separated
by a few millimeters. By correlating hits between the two sensors, and
making use of the bending in the CMS 3.8~T magnetic field, pairs 
of hits, referred to as stubs, with a small bend (high $\pt$) can be
selected. As illustrated in Fig.~\ref{fig:pTsketches}, 
this concept works both for barrel modules and modules in an
endcap geometry. As the hit separation in the azimuthal 
direction is a function of the
radial separation of the hits, the endcap geometry will require a
larger sensor separation at large pseudorapidity ($\eta$). 
The goal of the
$\pt$ modules is not to provide a precise momentum measurement, but to distinguish
stubs with a transverse momentum greater than 2~GeV, a threshold that corresponds to a data volume reduction of $\mathcal{O}(10)$, so that these can be propagated to the back-end track finding system. The most
challenging stub forming configuration is for the innermost barrel layer. At a radius of
about 250~mm, a 2~GeV track will have a separation $\Delta x$ of about 185~$\mu$m
for a sensor separation $\Delta r$ of 2.6~mm. With a pitch between the silicon strips of about 
100~$\mu$m, $\pt$ discrimination can be implemented to reject stubs
from low-$\pt$ trajectories. Though a larger sensor separation would
increase the hit separation in the two sensors and provide a more
precise $\pt$ determination, the larger stub formation windows would
increase the rate of combinatorial stubs. This is an issue in the
inner region of the detector where the occupancy is high. In the outer
part of the detector larger windows can be used since the
probability of forming stubs from uncorrelated hits is smaller.

\begin{figure}[tbp]
\begin{center}
\includegraphics[width=0.99\textwidth]{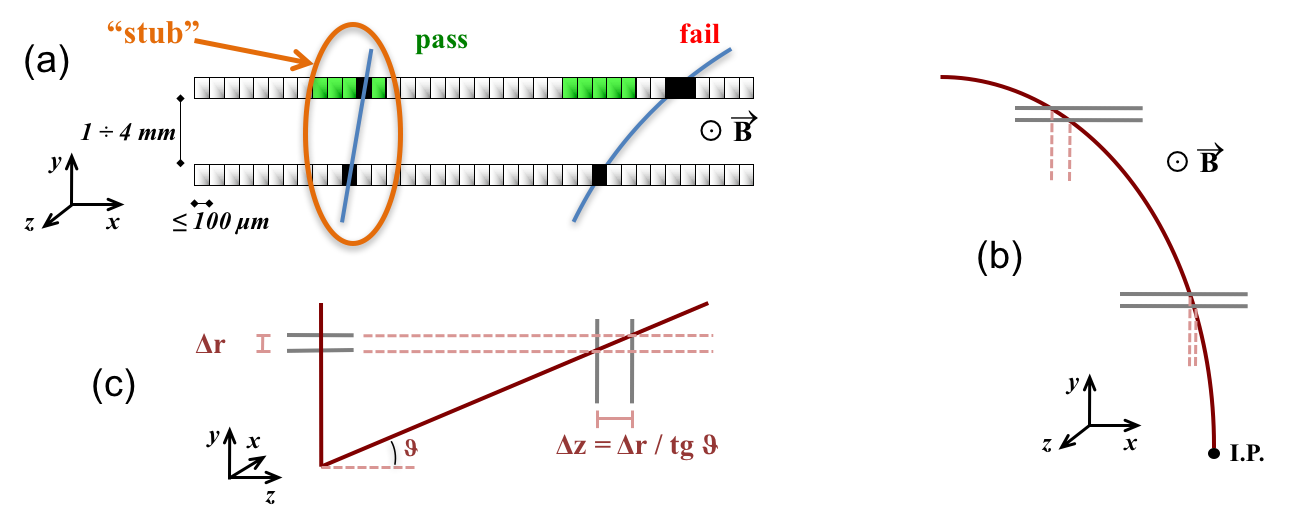}
\caption{\label{fig:pTsketches} The principle for $\pt$
  discrimination in the outer tracker modules.
(a) Correlation of signals in closely-spaced
sensors enables the rejection of low-$\pt$ particles; the channels in
green show the acceptance
window to form a stub from the hit indicated in the inner sensor. (b) The same transverse momentum corresponds to a
larger distance between the two hits at larger radii for a given sensor spacing. (c) For the endcap
discs, a larger spacing between the sensors is needed to achieve the same discriminating
power as in the barrel at the same radius. Figure adapted from Ref.~\cite{TrackerTDR}.
}
\end{center}
\end{figure}

The CMS tracker will make use of two types of $\pt$ modules. The first
type is the pixel-strip (PS)
module in which one sensor tier consists of macro-pixels (1.446~mm 
long by 100~$\mu$m) and the other tier of strips (2.4~cm long). The other module type is the
strip-strip (2S) module where both sensor tiers are strips (5~cm long
and 90 $\mu$m pitch). These
modules are illustrated in Fig.~\ref{fig:modules}. For both PS and 
2S modules, hits are read out at the edges of the module and
the data are communicated between the two sensor tiers 
through a flex circuit in order to correlate hits and produce stubs. 

Stubs are formed in the MacroPixel ASIC (MPA)~\cite{Ceresa:2019xcv} in the PS
modules and in the CMS Binary Chip (CBC)~\cite{Prydderch:2018pxp} in the 2S modules. 
The stubs formed in the front-end ASICs are communicated 
to the concentrator chips (CIC)~\cite{Nodari:2019dnm}. There are two CICs on
each module; one for each readout side. The CICs 
implement load balancing by grouping stubs from eight consecutive
bunch crossings before sending the payload to the back-end electronics
via optical links.
The CIC chips implement stub $\pt$ ordering
and discard the stubs with lowest $\pt$ in case 
truncation is needed. Readout data and module configuration data are
transmitted and received using one Low-power Gigabit Transceiver (LpGBT)~\cite{Mendez:2019bys} per module. Depending
on the location in the detector (and thus the occupancy) the optical links are operated at either 5 or 10~Gbits/s. 

The PS modules are used in the inner barrel layers and the inner half of the disks as they provide 3-D space points that allow precise $z$ position measurements for the 
reconstructed tracks. The PS modules are also capable of handling the
higher hit rate due to the finer segmentation with the macro-pixels and a smaller detector area.

\begin{figure}[tbp]
\begin{center}
\begin{minipage}[t]{0.52\textwidth}
\includegraphics[width=\textwidth]{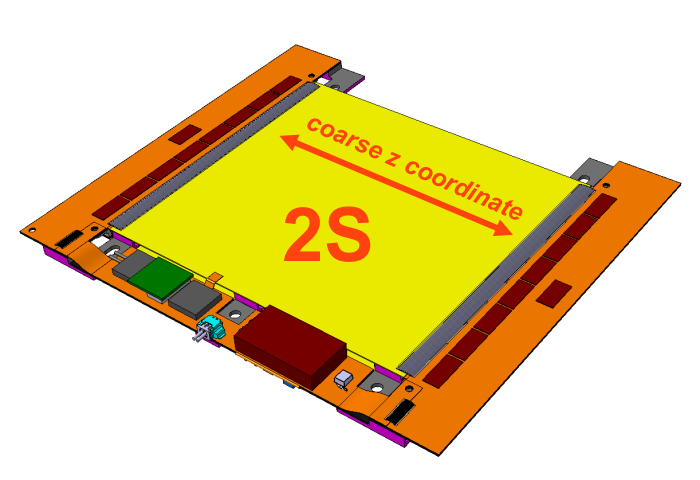}
\end{minipage}%
\begin{minipage}[t]{0.46\textwidth}
\includegraphics[width=\textwidth]{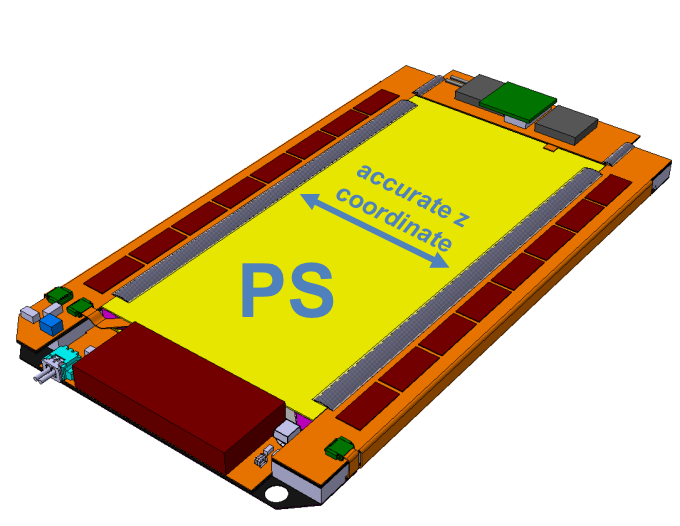}
\end{minipage}
\hfill 
\begin{minipage}[t]{0.52\textwidth}
\includegraphics[width=\textwidth]{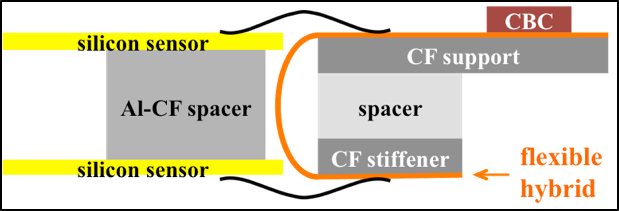}
\end{minipage}%
\hskip 2mm 
\begin{minipage}[t]{0.46\textwidth}
\includegraphics[width=\textwidth]{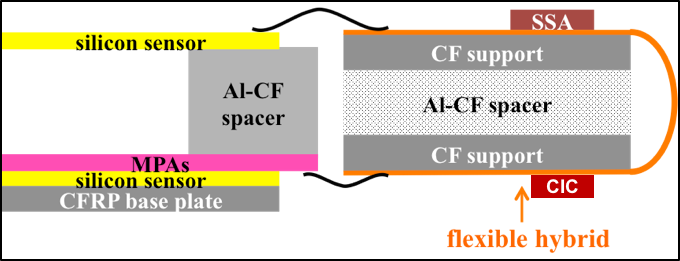}
\end{minipage}%
\caption{\label{fig:modules} 
Illustration of the concept of the $\pt$ modules for the upgraded 
CMS outer tracker for HL-LHC. The two types of modules, 2S and PS, are
shown to the left and right, respectively. The top images show a layout
of the two module types and the bottom images show a cross-sectional
view of the connectivity at the edges of the modules. These figures
illustrate how hit information is communicated between the two 
sensor tiers and correlations, stubs, are formed. In the 2S modules
one CBC reads out the hits from both sensors and forms the
correlations. For the PS modules, the strip sensor, at top in the
figure, is read out by the SSA and the hits are communicated through
the flexible hybrid to the MPA, which reads out the macro pixels and
form the stubs. The separation between the sensors varies from
1.6~mm to 4.0~mm. From Ref.~\cite{TrackerTDR}.
}
\end{center}
\end{figure}

The detector layout is shown for one quadrant of the detector in 
Fig.~\ref{fig:Tracker_Layout}. There are six barrel layers with the
inner three layers composed of PS modules (TBPS) and the outer 
three layers composed of 2S modules (TB2S). The modules in the
forward region of the
three inner layers are tilted such that charged particles from the interaction point (IP) will
traverse the modules in a direction approximately 
perpendicular to the sensor plane. This increases the efficiency
for reconstructing a stub since the particles are more likely to 
hit both sensors. It also reduces the sensor area needed to 
provide complete coverage. The PS modules in the TBPS use sensor
spacings of 1.6~mm, 2.6~mm, or 4.0~mm depending on the position
and orientation of the sensors, as shown 
in Fig.~\ref{fig:Tracker_Layout}. The 2S modules in the barrel
all have 1.8 mm spacing. 

There are five disks (TEDD) on each side of the interaction point.
Each disk has five outer rings of 2S modules with sensor 
spacings of 1.8~mm or 4.0~mm. 
The two disks closest to the IP extend somewhat closer to
the beamline and have ten rings of PS modules while the
outer three disks have seven rings of PS modules. 
Also shown in Fig.~\ref{fig:Tracker_Layout} is the stub acceptance
window, in number of strips,  for which hits in the 
two sensors are accepted as a stub. 
This window varies from as little as two strips in the PS
modules at the lowest radii in the forward region to nine strips.
For the 2S modules the acceptance window varies between 6--15
strips. These acceptance windows are configurable and can be
tuned to manage the rate for the trigger data.

\begin{figure}[tbp]
\begin{center}
\includegraphics[width=0.99\textwidth]{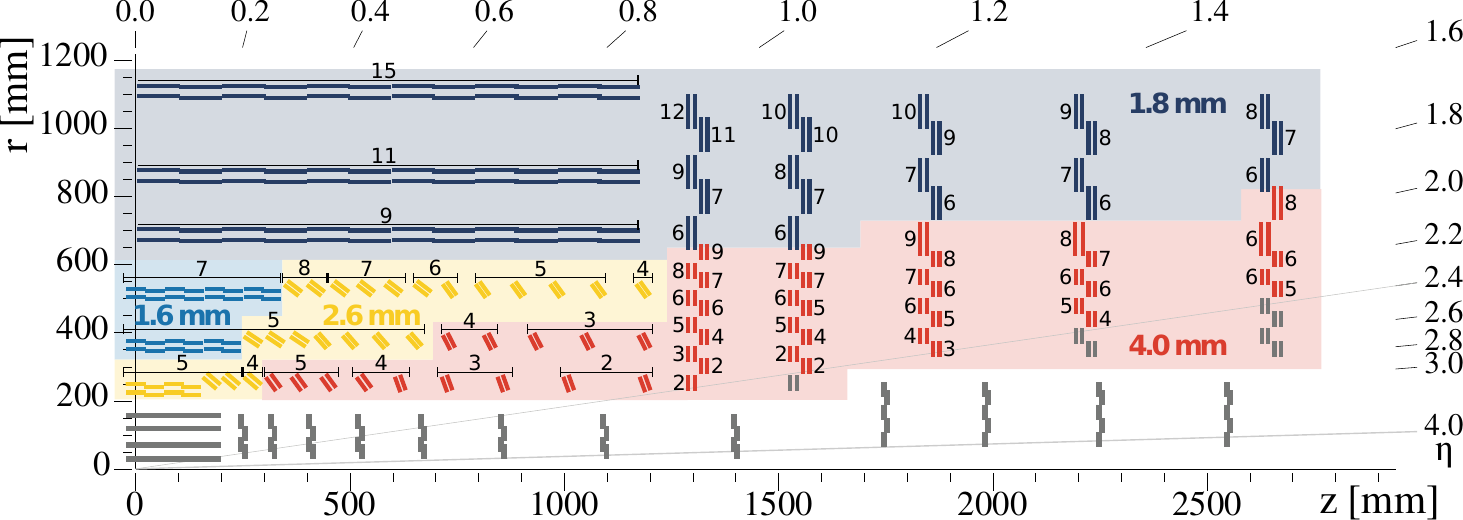}
\caption{\label{fig:Tracker_Layout} 
One quarter of the layout of the CMS outer tracker for HL-LHC, showing also 
the different module spacings and stub acceptance windows used. 
The PS modules are indicated in light blue, yellow, and red 
(the PS modules in grey are not used in the trigger). The different 
colors correspond to the sensor separation in the modules; blue is 
1.6~mm, yellow is 2.6~mm, and red is 4.0~mm. The 2S modules are in 
dark blue or red and have 1.8~mm or 4.0~mm sensor spacing, respectively. The numbers in black 
next to the modules are the stub acceptance windows in number of strips.
The inner pixel detector modules (grey) are not used in the L1
readout. From Ref.~\cite{TrackerTDR}.
}
\end{center}
\end{figure}

The simulated stub reconstruction efficiency as a function of particle $p_{\rm T}$ is shown in 
Fig.~\ref{fig:Stub_pt_turnon} for modules in the barrel and endcap regions.  
The stub finding windows are chosen to provide high efficiency at the 
2 GeV threshold for track finding. In the innermost layer, TBPS layer 
1, the turn-on curve for the stub finding efficiency is less sharp 
than in the outer layers due to the smaller bend of the track at smaller radii. 

\begin{figure}[tbp]
\begin{center}
\includegraphics[width=0.90\textwidth]{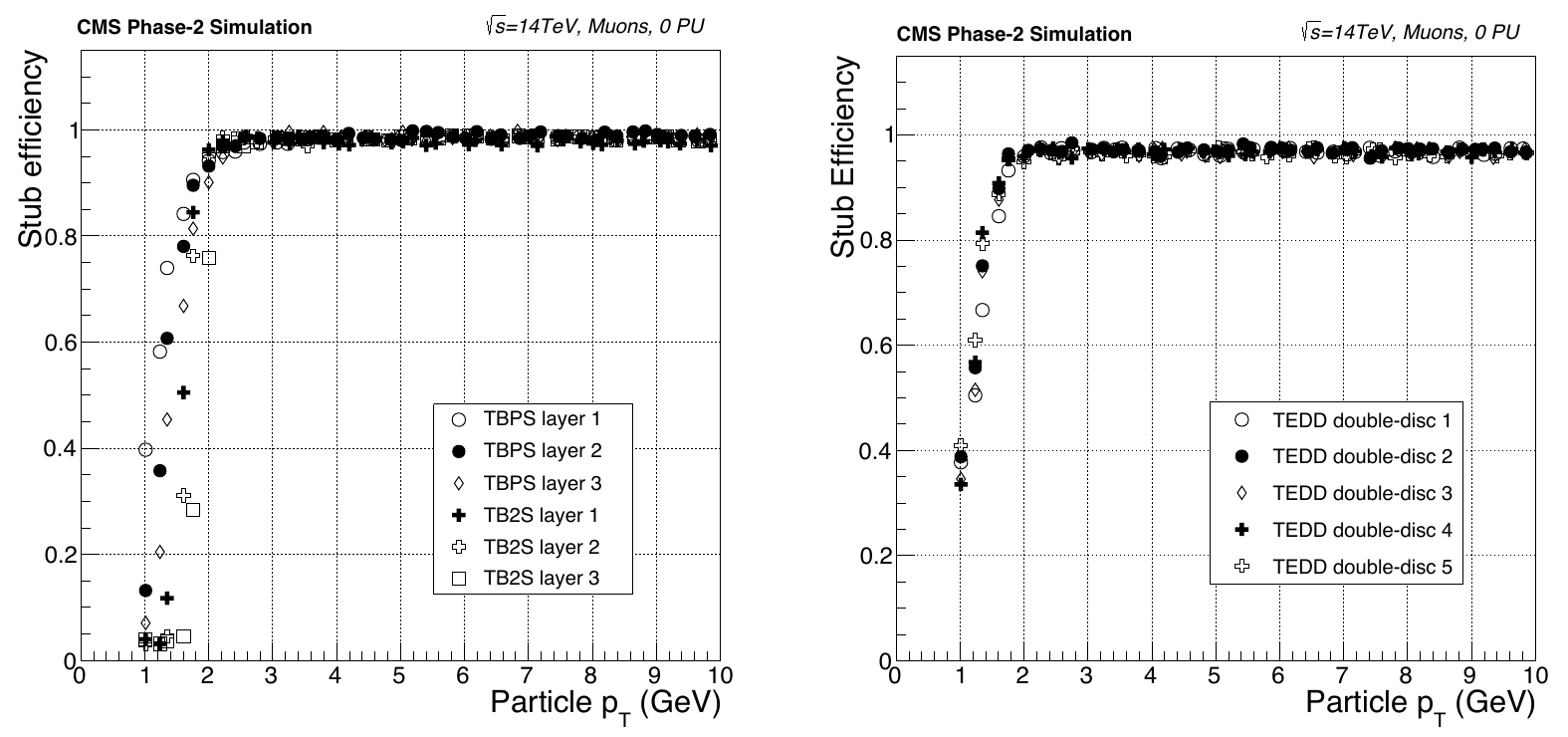}
\caption{\label{fig:Stub_pt_turnon} The stub finding efficiency as a function of particle $p_{\rm T}$ for muons, shown separately for the six barrel layers (left) and for the 5 disks (right). From Ref.~\cite{TrackerTDR}. 
}
\end{center}
\end{figure}

The full outer tracker design consists of 5,616 PS modules and 7,680 2S
modules for a total of 13,296 modules. 
Each module is read out and 
controlled using one LpGBT module. The off-detector readout and control is handled
by 216 DTC (Data, Trigger, and Control) boards. Each DTC is
capable of controlling up to 72 front-end modules. The DTC receives
data from the front-end modules and extracts the L1 accept data 
from the 40~MHz trigger data. The DTC unpacks the trigger data 
and assigns the stubs to the correct bunch crossing; it also implements
the data routing for the time and space multiplexing used by the
track finder boards. The optical data links from the
DTCs to the track finder boards operate at
25~Gbits/s.

%%%%%%%%%%%%%%%%%%%%%%%%%%%%%%%%%%%%%%%%%%%%%%%%%%%
\subsection{Trigger Architecture}

CMS utilizes a two-level trigger system to identify and select interesting collision events based on an initial L1 trigger, followed by a high-level trigger (HLT). The L1 trigger uses algorithms that run on custom electronics boards, while the HLT is implemented with algorithms that run on commercial CPUs. More recently, also algorithms running on GPUs are explored for the HLT.  
The L1 trigger has stringent constraints on the available processing time, referred to as the ”latency”, based on the available on-detector readout buffers that must temporarily store the information from different collision events until a signal is received as to whether or not to further process a given event. Strict constraints are also enforced on the total accepted event rate, following the design of the subsystem readout electronics. 
The CMS L1 trigger system for LHC Runs 1--3 has a latency of 4~$\mu$s
and a maximum readout rate of 100~kHz. For the HL-LHC operation, this will be increased to a latency of 12.5~$\mu$s and a maximum output rate of 750~kHz. The increased latency is necessary to accommodate the processing of L1 tracking~\cite{trigger_interim_tdr}. 

The entire CMS trigger system will be replaced for HL-LHC. At the core
of the redesigned system is the added capability of L1 tracking. 
Figure~\ref{fig:trigarch} shows a schematic overview of the upgraded CMS L1 trigger architecture.   
Barrel and endcap calorimeter trigger systems will process high-granularity information from the calorimeters. 
Barrel, endcap, and overlap muon track finding systems 
will provide triggering of muons up to $|\eta|<2.5$. The Global Track Trigger will reconstruct primary event vertices and define track-only based trigger objects. A two-stage {\it correlator} system will match L1 tracks with information from the calorimeter and muon systems, and perform a L1-adopted version of full event reconstruction (particle flow reconstruction)~\cite{PF_CMS}, to identify physics objects such as electrons, photons, muons, hadronic $\tau$ leptons, jets, and energy sums. This list of objects will be propagated to the global trigger, which decides whether to retain the event for further processing (an L1 accept) or to discard it.

\begin{figure}[tbp]
\begin{center}
\begin{minipage}[t]{0.9\textwidth}
\includegraphics[width=\textwidth]{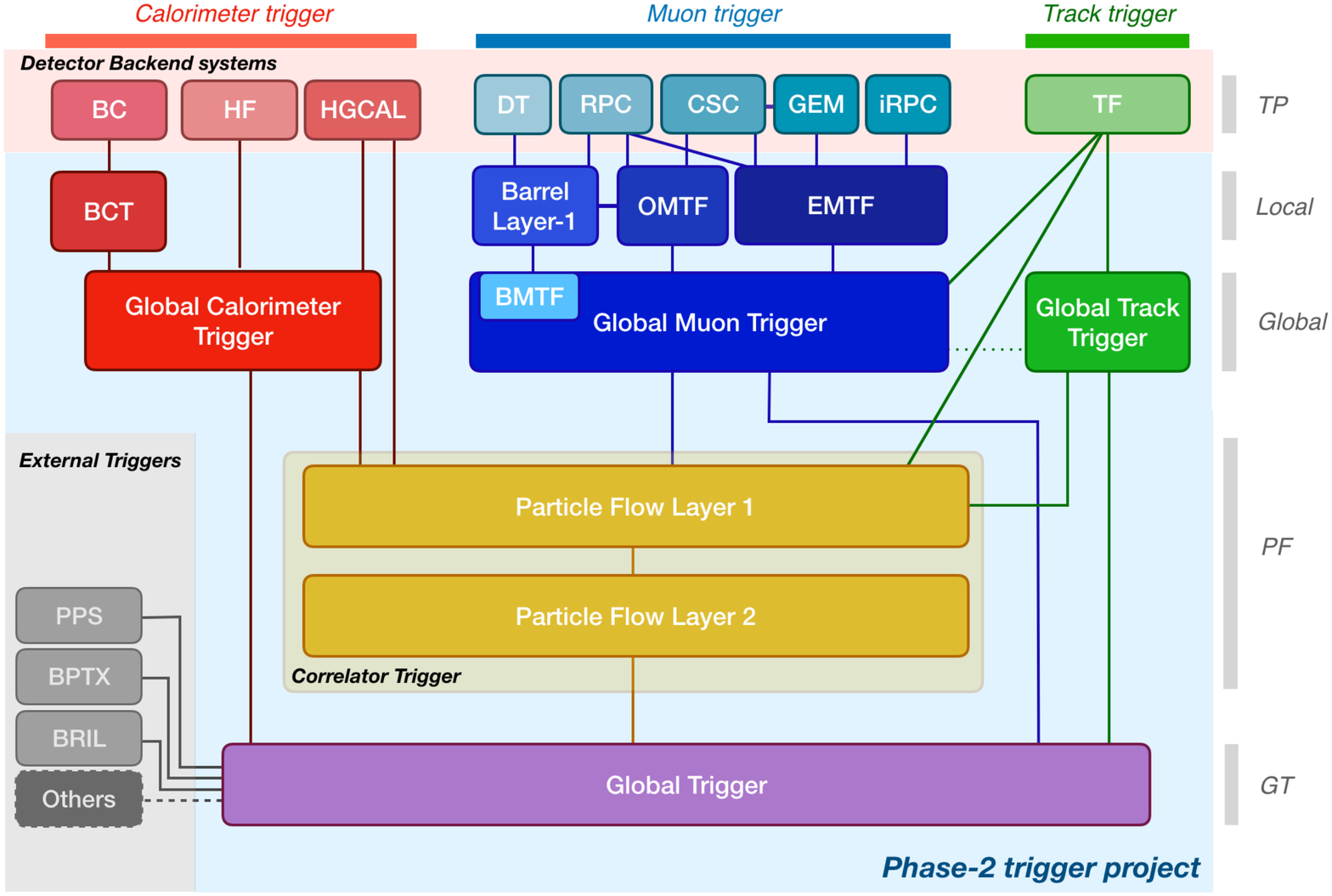}
\end{minipage}
\caption{\label{fig:trigarch} Overview of the CMS L1 trigger system for HL-LHC~\cite{CMS_P2_L1TDR}. The primary data flow is shown with solid lines; additional data paths are under study (dashes lines). The calorimeter trigger uses Trigger Primitives (TP) from the barrel calorimeter (BC), forward calorimeter (HF), and high-granularity calorimeter (HGCAL); it is composed of a Barrel Calorimeter Trigger (BCT) and a Global Calorimeter Trigger. The muon trigger receives inputs from the Drift Tubes (DT), Resistive Plate Chambers (RPC), Cathode Strip Chambers (CSC), and Gas Electron Multiplier chambers (GEM); it has a barrel layer-1 processor and muon track finders processing data from separate $\eta$ regions (BMTF, OMTF, and EMTF, respectively). The Track Finders (TF) provide L1 tracks to a Global Track Trigger (GTT) as well as other components of the system. A two-stage Correlator Trigger performs correlations between tracks and other objects, including a Level-1 adopted version of particle-flow (PF) reconstruction. The Global Trigger (GT) receives all L1 objects and issues the final trigger decision. 
}
\end{center}
\end{figure}

The inclusion of tracking is critical to maintain sufficiently low thresholds for electroweak physics and expanding the phase space of sensitivity for beyond-SM searches. 
Requiring e.g. multiples jets or leptons to originate from a common collision vertex (or $z$ position) reduces significantly the event rates due to accidental matches of objects from different overlapping $pp$ interactions. 
The definition of a complete L1 trigger menu is explored in
Ref.~\cite{trigger_interim_tdr}, which with the inclusion of L1
tracking gives a total rate of approximately 500~kHz (keeping a 50\%
safety factor with respect to the maximum output rate of 750~kHz);
without track information this would correspond to 4,000~kHz, a rate much too high to sustain. 
Incorporating tracking in the L1 trigger is thus essential in enabling the CMS HL-LHC physics program.

%%%%%%%%%%%%%%%%%%%%%%%%%%%%%%%%%%%%%%%%%%%%%%%%%%%
\subsection{L1 Track Reconstruction Approaches} 

As described in Sect.~\ref{sect:tracker}, the CMS outer tracker for
HL-LHC has been designed specifically to 
provide trigger primitives for L1 tracking. As part of validating 
the CMS tracker design, the implementation of the L1 tracking was 
investigated using different approaches, described in the Technical Design Report for the upgraded tracker~\cite{TrackerTDR}. 
Three distinct approaches were studied for performing the pattern recognition: the Tracklet approach, the Hough Transform approach, and the Associative Memory (AM) approach. 
The first two were fully based on FPGAs, while the latter used dedicated AM ASICs for pattern matching. 
Each pattern recognition approach was followed by a final track fit that improves track parameters and
selects the best stub combinations in the case of
combinatorics. The
work required by the track fit varies depending on the pattern
recognition approach; approaches that make use of coarse stub positions will have
additional combinatorics and fake patterns that must be 
rejected during the final track fit. For all approaches the track fit was implemented in FPGAs. 

The three pursued L1 tracking approaches have in common that they split the detector
into smaller geometrical regions and utilize time multiplexing
to distribute the tasks of performing the pattern recognition, track fitting, and 
duplicate removal to many different hardware components.
Specifically, data are organized in regions 
in $\eta-\phi$ (either in $\phi$ sectors spanning the full $\eta$
region, or in $\eta-\phi$ trigger towers), and organized 
using a round-robin time-multiplexing system with $n$ 
identical copies of the system. The choice of detector segmentation
and time-multiplexing factor depends on the algorithms and are
described below.
The described algorithms have been implemented in firmware, and 
validated using hardware demonstrators. 
For the demonstrators developed in support of Ref.~\cite{TrackerTDR},
the 
stub data pre-processing in the DTCs was not fully specified; the 
different approaches therefore made different assumptions of its functionality.

This section discusses the common technology enablers necessary for L1 tracking, followed by a description of the different
 track finding approaches along with their implementations.
The demonstrator systems and hardware platforms validating their feasibility are also described.

%%%%%%%%%%%%%%%%%%%%%%%%%%%%%%%%%%%%%%%%%%%%%%%%%%%
\subsubsection{Technology enablers}

The requirements for the HL-LHC track triggers in terms of
data volume and processing needs are significantly higher than
those of earlier track trigger implementations described in 
Sect.~\ref{sect:earlydevelopments}.
However, over the last decade there has been rapid development 
in several technologies that enable the implementation of 
a hardware track trigger for the HL-LHC. The primary challenges are 
the input data rate, the processing requirements to find the tracks, 
and the latency required in the L1 trigger.
The input data rate to the L1 track finding system is on the order of
30~Tbits/s, corresponding to about 15,000 stubs per bunch crossing
every 25~ns. Sufficient computational processing power is required to 
implement the pattern recognition and track fitting algorithms. These
algorithms have to process the roughly 15,000 stubs that arrive every 
25~ns and find the tracks within about 4~$\mu$s. 

The rapid development in the last decade has provided optical links
that operate at speeds of 25~Gbits/s. The use of these high speed 
links allows moving the data into the trigger system with a manageable 
number of links and with low latency. The rapid development of 
FPGAs provides the computational power required to implement the 
algorithms. The latest generation FPGAs have several thousands 
Digital Signal Processor (DSP) and block RAM units, and sufficient 
routing and logic support 
to implement the algorithms for the track finding
(\url{https://www.xilinx.com/support/documentation/selection-guides/ultrascale-plus-fpga-product-selection-guide.pdf}, or~\url{https://www.intel.com/content/dam/www/programmable/us/en/pdfs/literature/pt/intel-agilex-f-series-product-table.pdf}).

%%%%%%%%%%%%%%%%%%%%%%%%%%%%%%%%%%%%%%%%%%%%%%%%%%%
\subsubsection{Tracklet Approach}
\label{sec:tracklet}

The tracklet approach is an implementation of a traditional road
search pattern recognition algorithm. Seeds are formed from stubs in adjacent layers or 
disks and matching stubs are found in the other detector layers. A
final track fit is performed including the stubs matched to the seed. The 
tracklet approach makes use of the full stub position resolution 
when forming the seeds. This allows finding track candidates
with high purity, which reduces the resources needed in the final
track fit.

\paragraph{Tracklet road search algorithm}

Tracklets, or the seeds, are formed from pairs
of stubs in adjacent layers or disks. For each stub pair a
trajectory is calculated, using the beam spot as a constraint in the
transverse plane, and projections to other layers and disks computed.
Using the projections to other layers and disks, matching stubs are
found and used in the final track fit. These steps are illustrated in 
Fig.~\ref{fig:trackletseeding}. The seeding is performed, in parallel, in
multiple combinations of layers to ensure coverage and redundancy.

\begin{figure}[tbp]
\begin{center}
\begin{minipage}[t]{0.32\textwidth}
\includegraphics[width=\textwidth]{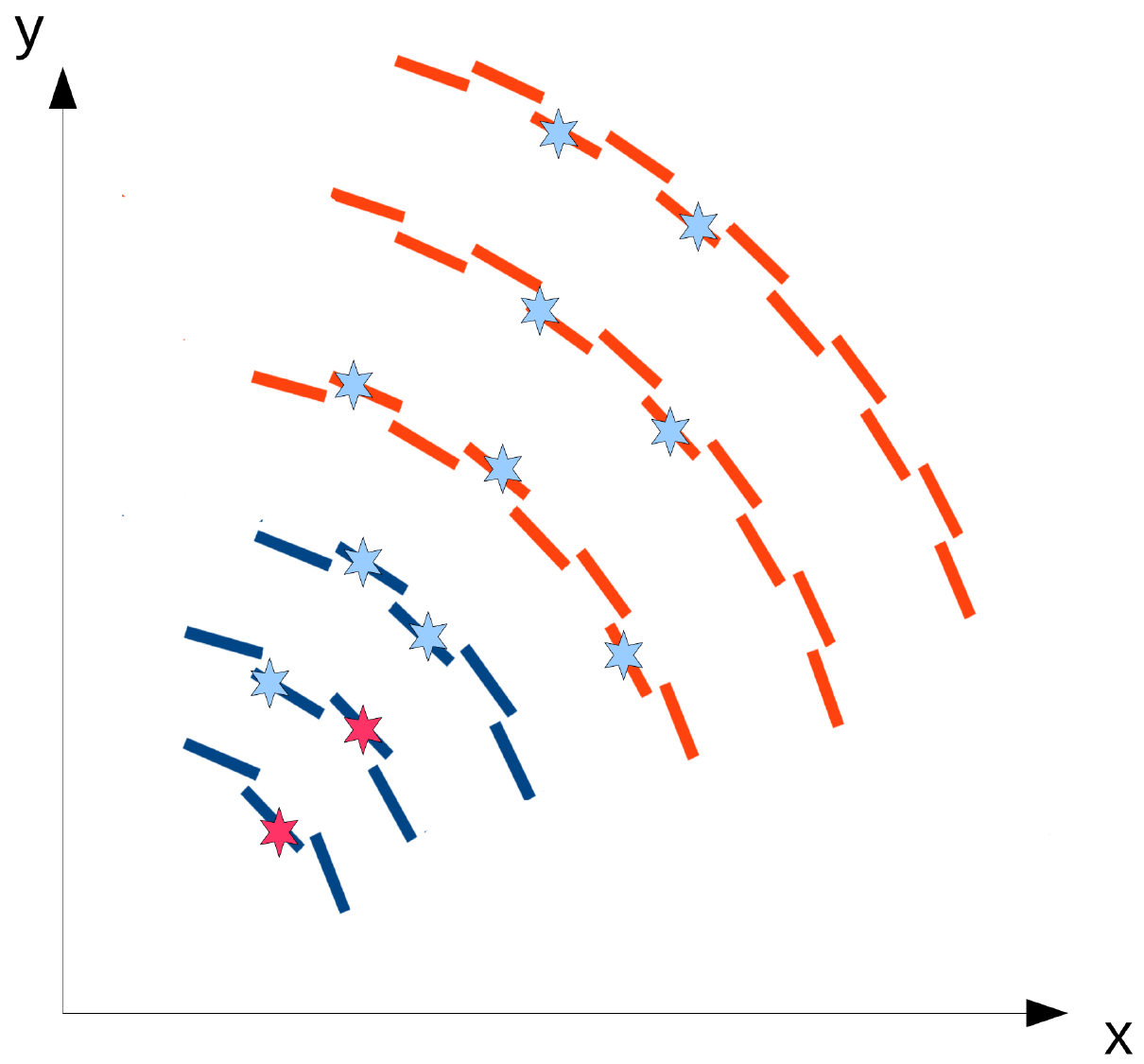}
\end{minipage}
\begin{minipage}[t]{0.32\textwidth}
\includegraphics[width=\textwidth]{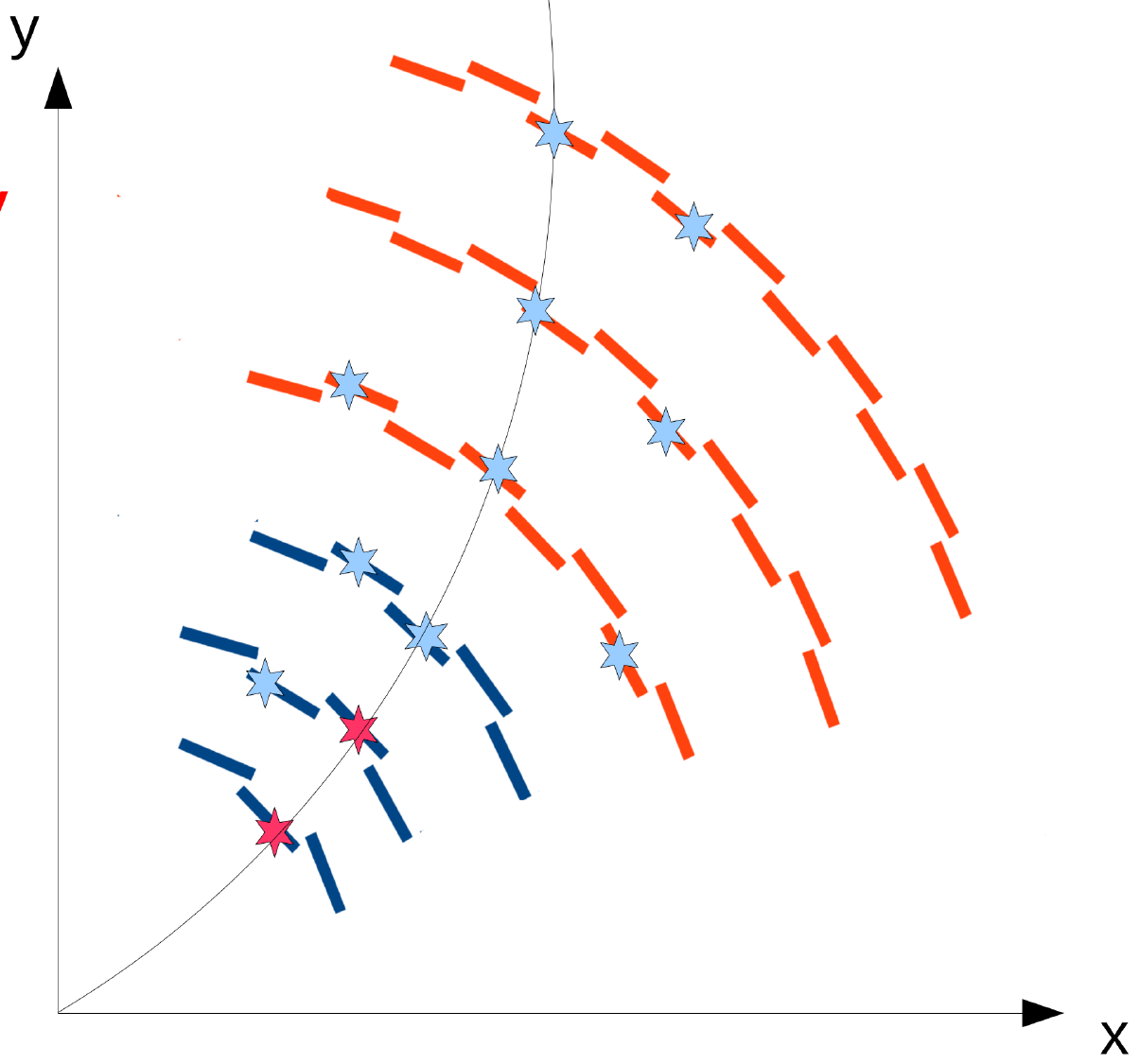}
\end{minipage}
\begin{minipage}[t]{0.32\textwidth}
\includegraphics[width=\textwidth]{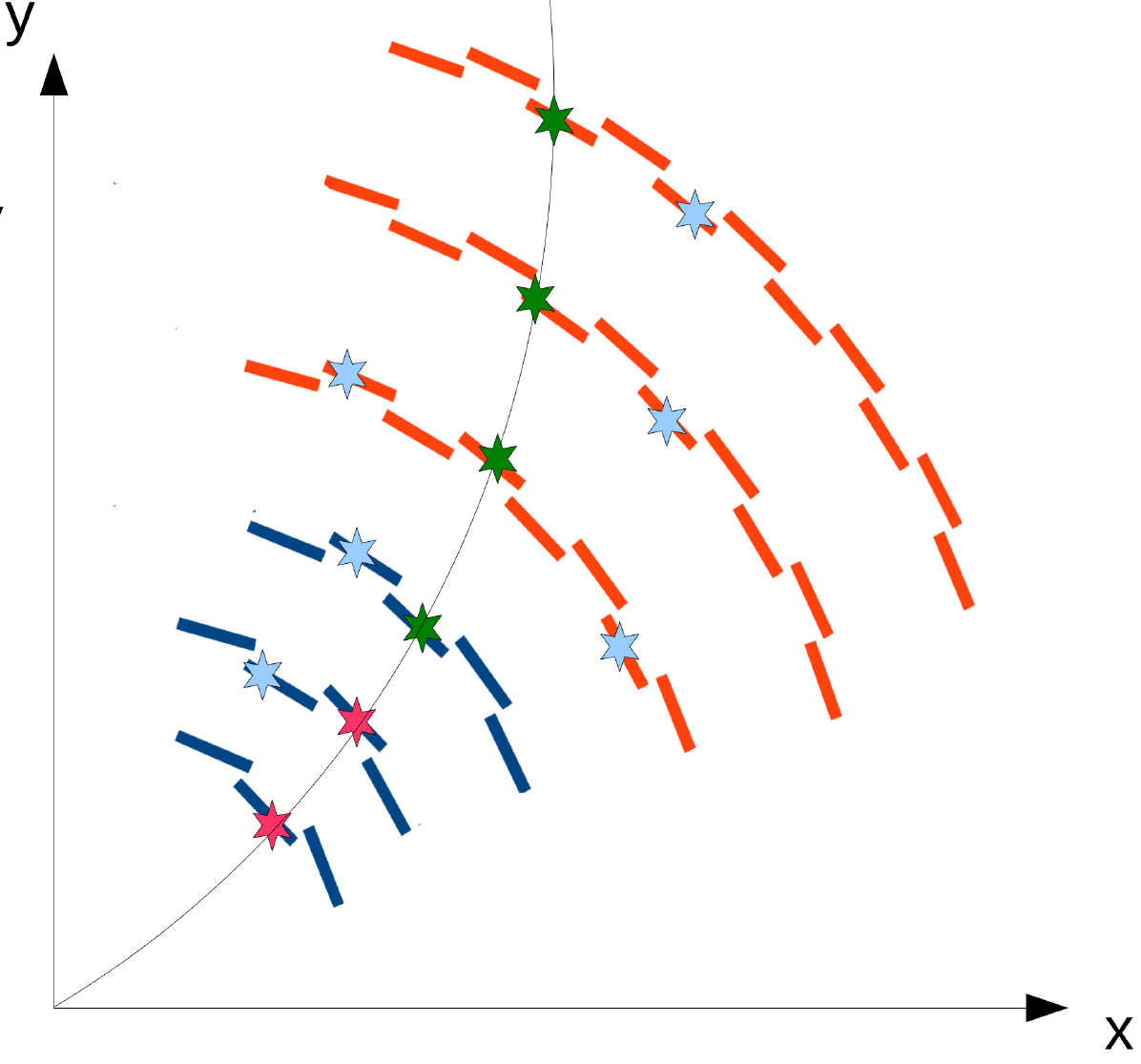}
\end{minipage}
\caption{\label{fig:trackletseeding} The tracklet approach forms seeds
from pairs of stubs in adjacent layers or disks (left). The seed trajectories are projected to other layers/disks (middle) where stubs are matched and fit to form the final track (right). From Ref.~\cite{TrackerTDR}.
}
\end{center}
\end{figure}

The main challenge in implementing this approach lies in organizing
the data such that sufficient processing capacity is available and
truncation in the fixed latency environment is acceptable. This
is achieved by dividing the data within each sector
into finer regions, referred to as {\it virtual modules}.
By forming stub pairs by combining stubs in pairs of virtual modules, 
the number of pairs that are not consistent with a real seed can
be significantly reduced.

There are four main steps in the implementation of the tracklet
approach; the stub organization, the tracklet formation,
the stub matching, and the track fit. The first step organizes the 
stubs into regions based on the stub $\phi$ position and the stub 
$z$ or $r$ position in the barrel and disks, respectively. This
organization of the stubs allows for the tracklet finding to proceed
by first forming candidate pairs of stubs. These stub pairs are 
required to be consistent with $\pt>2$~GeV and
$|z_0|<15$~cm, and the bend of the two stubs must be consistent with the tracklet $\pt$.
 
Stub pairs that are selected are sent to the tracklet calculator
where the precise track parameters as well as projections to other
layers and disks are calculated. These projections are calculated with respect to a nominal 
layer or disk position, 
and the derivatives of the $\phi$ and $r$ ($z$) positions are
also evaluated. Stub matches in the 
other layers and disks are first looked for coarsely using the 
projection to the nominal radius; in a second stage the
precise residuals are calculated using the exact stub 
position. The
implementation selects the best stub match in each layer or disk based on the $\phi$ 
residual.

\paragraph{Linearized $\chi^2$ track fit}

For the implementation of the final track fit in the tracklet approach, 
a linearized $\chi^2$ fit is used that takes advantage of the 
information from the pattern recognition step. To match stubs to the projection from the seed, the residuals between
the projections and stubs are calculated in both $\phi$ and $z$ (or
$r$ for disks). Using these residuals, $\delta y_i$, and linearizing
the $\chi^2$ fit, we can express the final track parameters 
$u=(\rho^{-1}, \phi_0, t, z_0)$ where $\rho$ is the signed radius of
curvature related to $q/\pt$, $\phi_0$ is the track azimuthal angle at
the IP, $t=\sinh\eta$, and $z_0$ is the longitudinal impact parameter as
\begin{equation}
u = \bar u + \sum_iM_i\delta y_i
\end{equation}
where $\bar u$ are the track parameters from the seed and 
$M$ is a weight matrix. The weight matrix can be precomputed,
as to a good approximation it is independent of the parameters
except for stubs in the disks where it depends on the parameter
$t=\sinh \eta$ and the weights are tabulated for different ranges of $t$.
The linear form of this equation means that 
the updated track parameters are obtained simply by multiply-and-add 
operations that are efficiently implemented in hardware. 

Tracklet seeds with two or more matches in
other layers or disks have a final track fit performed. These
tracks have a minimum of four stubs. Since the track
parameters from the tracklet seeds are accurately calculated based
on the precise positions of the stubs, the final track fit provides
a refinement to the track parameters. A linearized $\chi^2$ fit
that updates the track parameters of the seeds therefore works well.

\paragraph{Tracklet demonstrator implementation}

The tracklet algorithm implementation relies on extensive 
parallelization, both in time and space. For the hardware 
demonstration of the algorithm the detector was 
divided into 28 $\phi$ sectors.
Each time slice in a sector is processed by a 
{\it sector processor}, which is a dedicated hardware board with
a modern FPGA with sufficient resources for the algorithm 
implementation. 
For the demonstrator implementation, a time-multiplexing factor of six 
was assumed, corresponding to each sector processor receiving a 
new event every 150~ns. 

The algorithm is implemented in eight 
processing steps: 1) the VMRouter organizes the stubs
into the virtual memories for the tracklet finding, 2) the 
TrackletEngine forms candidate stub pairs, 3) the 
TrackletCalculator calculates the precise trajectories
and projections, 4) the ProjectionRouter organizes the 
projections into virtual modules, 5) the MatchEngine
forms projection stub candidate matches, 6) the
MatchCalculator calculates the precise residuals
between the projections and stubs, 7) the 
FitTrack modules perform the final track fit, and
8) the PurgeDuplicate removes duplicate tracks. 
Two additional processing steps were used 
for matching stubs to projections pointing outside the
sector where the tracklet was formed. 
The functionality of each of these modules were implemented
in firmware (Verilog) and the complete project 
was built from connecting up many of these modules.
The calculations, primarily done in the TrackletCalculator,
MatchCalculator, and FitTrack modules, are implemented
using the DSP blocks in the FPGAs. The other processing 
modules primarily organize the data by routing data to the
relevant memories that act as buffers between the processing
steps. The implementation is pipelined such that a 
different bunch crossing is processed at the same time in
each of the eight processing steps. 

The tracklet algorithm was implemented using the Calorimeter
Trigger Processor (CTP7) $\mu$TCA board developed for the CMS Phase-1
upgrade~\cite{CTP7}. The CTP7 board has a Virtex-7 XC7VX690T FPGA and 63 input
and 48 output optical links operating at speeds up to 10~Gbits/s. The
demonstrator system implemented one time slice for one central sector and its
two neighboring sectors. 
A separate board was used as the source of input stubs
to the track finder boards and as the receiver of the final
tracks~\cite{tracklet_paper}.

%%%%%%%%%%%%%%%%%%%%%%%%%%%%%%%%%%%%%%%%%%%%%%%%%%%
\subsubsection{Hough Transform + Kalman Filter Approach}
\label{sec:HT}

The Hough transform is a common tool for pattern
recognition~\cite{HoughTransform} 
and is implemented in the track finding
to identify curved trajectories in the $r-\phi$ plane. In the
approach studied for Ref.~\cite{TrackerTDR}, the hit patterns from the Hough transform are processed by a Kalman filter as
the final track fit~\cite{tmtt_demo_paper}.

\paragraph{Hough transform pattern recognition}

The trajectory of
a charged particle produced at the origin (interaction point) that
travels through a uniform magnetic field, $B$,  satisfies
\begin{equation}
\phi_0=\phi-\frac{qB}{2\pt}\cdot r  
\end{equation}   
where $r$ and $\phi$ are the trajectory coordinates, $\phi_0$ is
the angle of the trajectory at the production point, $q$ is the
particle's electric charge, and $\pt$ the
transverse momentum. 

To reduce the correlation between the $r$ and $\phi$ positions it is
convenient to use $r_{\rm c}=r-C$, where $C$ is a radius 
approximately in the middle of the detector. This gives
\begin{equation}
\phi_{{\rm c}}=\phi-\frac{qB}{2p_{\rm T}}\cdot r_{\rm c} 
\label{eq:hteq}
\end{equation}  
where $r_{\rm c}$ is a signed radial position
and $\phi_{{\rm c}}$ is the trajectory 
angle at the radius $C$. 
For each stub, at a given $r_{\rm c}$ and
$\phi$, Eq.~\ref{eq:hteq} describes a straight line in
the $\phi_{{\rm c}}$ {\it vs.} $q/p_{\rm T}$ plane. For a track with
six stubs, as shown in Fig.~\ref{fig:htidea}, each of these
stubs will form a unique line, where the slope is a function of
the radial position of the stub. If the stubs belong to the same
particle trajectory these lines will all go through the same $(q/p_{\rm T},
\phi_{{\rm c}})$ point. Tracks can therefore be identified by looking for
points in the $(q/p_{\rm T}, \phi_{{\rm c}})$ plane where multiple
lines overlap. In an FPGA this is implemented using an array to
find bins with multiple entries.
This gives the track candidates with an estimate
of the track parameters in the $r-\phi$ plane.

\begin{figure}[tbp]
\begin{center}
\includegraphics[width=0.75\textwidth]{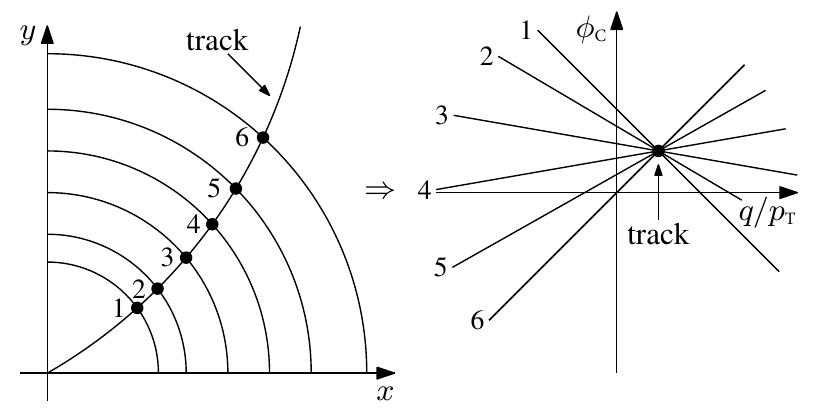}
\caption{\label{fig:htidea} Illustration of the Hough Transform idea
  for finding particle trajectories. The left-hand side shows a
  particle trajectory with six stubs in different layers. The right-hand
side shows the lines of possible $\phi_{\rm c}$ and $q/\pt$ for each of the
six stubs. The point where these six lines cross corresponds to the 
track parameters for the trajectory. Adopted from Ref.~\cite{tmtt_demo_paper}}
\end{center}
\end{figure}

\paragraph{Kalman filter track fit}

The Hough transform pattern recognition produces track candidates
that can have multiple stubs in a given layer and the final fit must therefore
filter the stubs used to form the track. 
This filtering and fitting is implemented using a Kalman 
filter~\cite{,Billoir:1990we}. 
The Kalman filter begins with an initial seed estimate of the track parameters and their uncertainties, referred to as a {\it Kalman state}. Here, the seed is provided from the Hough transform. 
The Kalman state is updated, iteratively, starting from stubs in the innermost layers. 
The seed trajectory is projected to the next layer or disk and
the state is duplicated if there are more than one stub. The track 
parameters are updated using the position information from the next stub. 
The fit allows for one missing layer.  This process is repeated until 
four stubs has been added to the trajectory. The pattern
recognition (Hough transform) requires at least five stubs, while a successful fit
requires four stubs.

\paragraph{Hough transform + Kalman filter demonstrator implementation}

The Hough transform + Kalman filter approach is implemented
as a Track Finding Processor with four distinct components: (i)
Geometric Processor (GP), responsible for the pre-processing of stub
data; (ii) Hough Transform (HT), coarsely
grouping stubs consistent with high-$\pt$ trajectories; (iii) Kalman
Filter (KF), second stage track cleaning and track fitting; (iv)
Duplicate Removal (DR), removing duplicate tracks. This approach has been implemented in firmware on FPGAs. The track finding processor handles $1/8$ of the detector, an octant, and uses a time-multiplexing factor of 36.

The first step involving the Geometric Processor carries out an initial processing of the stub data as it is assumed to be received from the DTCs into an extended 64-bit format, with motivation to minimize the downstream logic requirements in the Hough Transform. It also assigns stubs to 36 sub-sectors ($2 \phi \times 18 \eta$). 

Each Track Finder Processor uses 36 HT arrays running in parallel, each processing stub data consistent with the corresponding geometric region defined by the Geometric Processor; one GP sub-sector corresponds to one HT array. The HT array is split into two pipelined stages, first the filling of the array with stubs, and second the readout of the found track candidates. Each of these two stages processed one stub at 240~MHz. 
The HT array is implemented in firmware as 32 {\it columns} in $|q/\pt|$ and 64 {\it rows} in $\phi$. 

The Kalman filter was implemented in two parts, a data-flow component, which carries out the matrix operations described by the Kalman formalism (state updater and the calculation of track parameters and covariance matrix), as well as a control-flow component, which manages the stub and state data.

The Hough transform + Kalman filter approach was
implemented using the Master Processor 7 (MP7) $\mu$TCA
boards developed
for the CMS Phase-1 trigger upgrade~\cite{MP7}. 
The MP7 board uses the same Virtex-7 FPGA as the CTP7 
board. The demonstrator
system used five MP7 boards to implement the track reconstruction for one time slice for one
detector octant. 
One MP7 board was used as the Geometrical Processor, two
boards were used for the HT pattern recognition, and
two boards for the KF track fitter. In addition, two MP7 boards
were used as data sources and one was used as a sink for the produced 
tracks~\cite{tmtt_demo_paper}.

%%%%%%%%%%%%%%%%%%%%%%%%%%%%%%%%%%%%%%%%%%%%%%%%%%%
\subsubsection{Associative Memory + FPGA Approach}
\label{sec:AM}

The associative memory approach uses a content addressable memory
(CAM) ASIC to implement the pattern recognition. The AMs
allow simultaneous matching to a large number of patterns,
providing a fast response once all hits have been loaded.
The patterns are coarse hit positions, super strips, on the
modules. The AM returns all roads, i.e. patterns matched that have a
minimum number of matched stubs. The final $\chi^2$ track fit
implements filtering to reject wrong stub combinations.

\paragraph{Associative memory pattern recognition}
\label{sec:AMPR}

In the AM approach the detector is divided into trigger towers. The demonstration
system explored for Ref.~\cite{TrackerTDR} had 48 regions (8
in $\phi$ and 6 in $\eta$), and a time-multiplexing factor of 20.  Based on the demonstration studies, about 1M patterns were determined to be required for each trigger tower and 
time-multiplexing slice.  The design assumed that individual AM
chips could be produced with about 150k patterns.
The stub data received from the front-end
is routed by the Data Organizers (DO) to the appropriate processing unit that is handling the stub for that BX. The
stubs are converted to super strips and transmitted to the AM. The
super strips vary in width from about 1~mm at the inner radii up to
about 10~mm  at outer radii. The larger strip size at outer radii
optimizes the number of patterns required; larger strips at outer 
radii compensates for multiple scattering and lower occupancy. Once
all stubs are loaded into the AM the pattern recognition is performed. 
Stubs that are on roads found by the AM are passed to the Combination 
Builder that forms the different stub combinations from stubs within
each road. These are then passed onward to the track
fitter stage.

\paragraph{Track fitting for the AM approach}
\label{sec:AMTrackFit}

For the AM approach, a somewhat different implementation of a 
linearized $\chi^2$-fit is used, which is based on 
a principal component analysis (PCA) 
technique~\cite{AM_fitter_paper}. This fit obtains the track parameters by 
multiplying a vector of stub coordinates 
by an appropriate translation matrix. Again, the matrix 
multiplication can be efficiently implemented with low latency 
using the DSP resources in modern FPGAs. Tabulating the transformation 
matrix is very simple for a set of hits in a cylindrical geometry, 
while in a geometry with disks or varying radial positions within a 
layer, the matrix depends on the track parameters, particularly $\eta$. 
The fit used has implemented a method to project the hit positions to 
fixed radial positions.

\paragraph{Associative memory + FPGA demonstrator implementation}

The AM+FPGA approach was 
implemented using AM chips that perform the stub selection and 
pattern recognition using coarse stub position information, 
followed by track fitting implemented in an FPGA using 
stubs at full resolution. 
For each of the 48 trigger towers, the AM chips are preloaded with hit 
patterns representing possible valid trajectories for that tower's 
geometry. 
In the first processing step, 
stubs are received and routed to the pattern recognition 
mezzanine board for the relevant trigger tower and time-multiplexing slice. The data organizer formats and
loads the data into the AM. When all data has been loaded the
AM pattern recognition is performed and the matched 
patterns are read out. The DO retrieves the stubs on the matched
patterns and forward them to the track fit. 
 
The demonstration of the AM+FPGA approach is based on the
Pulsar2b ATCA board~\cite{pulsar2b}. The Pulsar2b board uses the
Xilinx Virtex-7 690T FPGA and allows board-to-board communication
across the ATCA backplane. The backplane communication is used in the
demonstrator to implement the time multiplexing. In the demonstrator 
the Pulsar2b served both as the data source
and the Pattern Recognition Board (PRB). The Pulsar2b supports two 
Pattern Recognition Mezzanine (PRM) boards that implements the pattern
recognition using AMs. For the demonstrator two 
types of PRMs are used. One is based on the AM06 ASIC~\cite{Annovi_2017} and
provided the pattern density required. However, as this AM ASIC was 
not developed for a L1 application it did not meet the latency
or data throughput requirements. The second PRM 
was developed for AM ASICs that were under development at the time of the
demonstrator. The PRM was developed such that it could support an
FPGA instead of the AM ASIC. 
For the demonstrator, the PRM with the FPGA option is used; the FPGA provide a clock cycle accurate
proxy for the AM ASIC. This
allowed the demonstration to meet the latency 
requirement but did not have the pattern density that is ultimately
required, but allowed demonstration of the rest of the system.

%%%%%%%%%%%%%%%%%%%%%%%%%%%%%%%%%%%%%%%%%%%%%%%%%%%
\subsection{HL-LHC Track-Finding System}

CMS has designed a novel tracker based on the concept of $\pt$ modules
that will provide stub primitives at 40~MHz for the L1 tracking. The
ability to reconstruct tracks in hardware at the L1 trigger level was
demonstrated using three different approaches. Following these
demonstrations, CMS has adopted a hybrid approach that makes use of the tracklet approach for the pattern
recognition and the Kalman filter for the final track fit. This
choice allows the implementation of the full algorithm in commercial FPGAs
without the need to develop custom ASICs for the pattern recognition.

\subsubsection{Hybrid implementation}

The three described approaches for pattern recognition and track
fitting were pursued to demonstrate the feasibility of L1 tracking at 40~MHz in the
HL-LHC environment with an average of 200 pileup interactions. 
The conclusion of the studies was that all three methods were feasible
and could implement L1 tracking with comparable performance and
meeting the latency requirement of about 4~$\mu$s. 
For the ultimate HL-LHC system, CMS has decided to pursue an all-FPGA based approach. The primary reason for this choice is~\cite{TrackerTDR} to reduce the risks associated with the development of the AM ASIC, which involves new technologies such as 28~nm or 3D 
integration. 

Following the choice of an all-FPGA solution, CMS is pursuing a hybrid implementation. The tracklet algorithm is used for
finding the candidate tracks, i.e. seeds are found and matched to stubs in other layers 
and disks, while the final track fit is performed using the Kalman filter. The combination 
of the tracklet approach with precise seeds and the iterative Kalman filter provides optimal performance. 
A schematic overview of the data flow from the outer tracker through the DTC boards and onward through the track finder processing is shown in Fig.~\ref{fig:hybrid}.
The hybrid system assumes a time-multiplexing factor of 18 and a division in $\phi$ into nine "nonants". 

\begin{figure}[tbp]
\begin{center}
\includegraphics[width=0.99\textwidth]{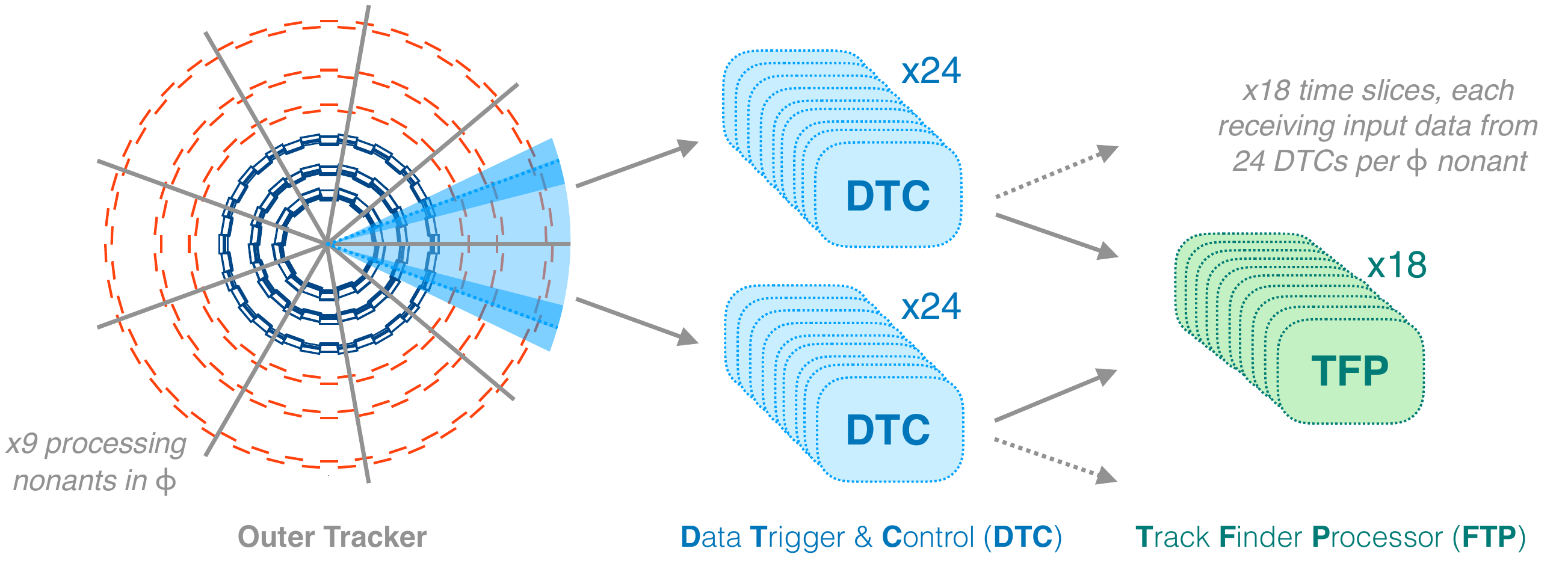}
\caption{\label{fig:hybrid} Schematic overview of the data flow from the outer tracker, divided into nine "nonants" in $\phi$, each read out by 24 DTC boards. From the DTCs, the stub data is forwarded to the track finder processor boards. 
}
\end{center}
\end{figure}

%%%%%%%%%%%%%%%%%%%%%%%%%%%%%%%%%%%%%%%%%%%%%%%%%%%
\subsubsection{Hardware platforms} 

Developments of hardware prototypes and plans for the ultimate system are underway to define the track finder boards that will be used for the HL-LHC system. 
The Apollo~\cite{APOLLO} platform will be used for the track finder processing boards, while the Serenity~\cite{Serenity} platform will be used for the DTC boards. Both platforms are based on the ATCA standard, and contain separate components for providing the necessary services and performing the data processing. The Apollo has a {\it Service Module} (SM) that provides infrastructure components, including the required ATCA Intelligent Platform Management Controller (IPMC)~\cite{IPMC}, powering, clock, and a system-on-module computer. A {\it Command Module} (CM) can be customized for the particular application, here the track finder processing, and contains two large FPGAs, several hundred optical fiber interfaces supporting link speeds of up to 28~Gbits/s, and memories. Similarly, the Serenity platform consists of a carrier card that provides the required services, such as powering, clocking, optical interfaces, interconnections between FPGAs, IPMC functionality, and an on-board CPU for controlling the board. Daughter cards host FPGAs that are responsible for the data processing, in this case, the DTC functionality.

%%%%%%%%%%%%%%%%%%%%%%%%%%%%%%%%%%%%%%%%%%%%%%%%%%%
\subsubsection{Displaced tracking}

The CMS L1 track finding approaches discussed so far are developed for
reconstructing prompt particles. However, the unique design of the CMS
Outer Tracker for HL-LHC also has the potential of reconstructing
long-lived charged particles that may have escaped detection thus far. Long-lived charged particles that decay a macroscopic distance away from the primary interaction point but prior to traversing the outer tracker would appear as {\it displaced} trajectories. With the vast physics motivation to search for such particles, see e.g. Ref.~\cite{longlived1} for a comprehensive review, efforts are underway to explore the full potential and technical feasibility of extending the CMS L1 tracking to reconstruct displaced trajectories. Proposed initially in Ref.~\cite{yuri1} and further explored in Ref.~\cite{yuri2}, the CMS L1 tracking could identify trajectories with displacements in the plane transverse to the beam line as large as about 10~cm. 
In particular, the physics case of a Higgs boson decaying to two new
light scalars that in turn decay to jets, where the scalars  have a
sufficiently large lifetime so that the jets appear displaced from the
primary vertex, is studied~\cite{YR_Higgs, YR_Exotic}. This process is
nearly free from Standard Model background contributions, and with the
current CMS detector would fail to be selected in the L1 trigger, while when incorporating displaced tracking, such rare exotic decays could be probed using the large HL-LHC data set. 
The dedicated reconstruction of displaced trajectories could be incorporated in the tracklet pattern recognition algorithm through the addition of {\it triplet} seeds that remove the constraint to the beamspot and instead adds a third stub~\cite{yuri1, james2019level1}.

%%%%%%%%%%%%%%%%%%%%%%%%%%%%%%%%%%%%%%%%%%%%%%%%%%%
\subsubsection{Expected performance} 

The expected performance of the L1 track reconstruction has been
studied using simulated data and validated with hardware demonstration
systems. The performance metrics used include the identification
efficiency of different types of charged particles, such as isolated
muons or electrons, and charged particles from the decays of top
quark--anti-quark pairs ($\ttbar$), which include charged particles
produced in the dense environment of jets. Other performance metrics
are the expected track parameter resolutions ($\pt$, $\phi_0$,
$\eta$, $z_0$), and the total rate of tracks per event.

Examples of the expected L1 tracking performance are illustrated in Fig.~\ref{fig:L1TK_perf}. It shows the expected L1 track reconstruction efficiency (top left) as a function of pseudorapidity for tracks in $\ttbar$ events overlaid with an average of 200 pileup interactions, using the hybrid algorithm~\cite{james2019level1}, as well as the expected $z_0$ resolution, corresponding to intervals that encompass either 68\% or 90\% of all tracks with $\pt > 2$~GeV (top right). At central $\eta$, the resolution is about 1~mm, while it is less precise at higher $\eta$ as a consequence of the CMS outer tracker geometry with tilted PS modules. 
Also shown (bottom left) is the expected track rate per $\phi$ sector (one ninth of the detector) for a minimum $\pt$ threshold of 2~GeV. 
Finally, the possible extension to reconstructed displaced trajectories that do not originate from the origin, consistent with possible long-lived particles, is shown (bottom right) as the efficiency as a function of the transverse impact parameter ($d_0$) for displaced muons in events without pileup. This extended, displaced tracking has the potential to significantly extend the $d_0$ coverage.

\begin{figure}[h]
\begin{center}
\begin{minipage}[t]{0.49\textwidth}
\includegraphics[width=\textwidth]{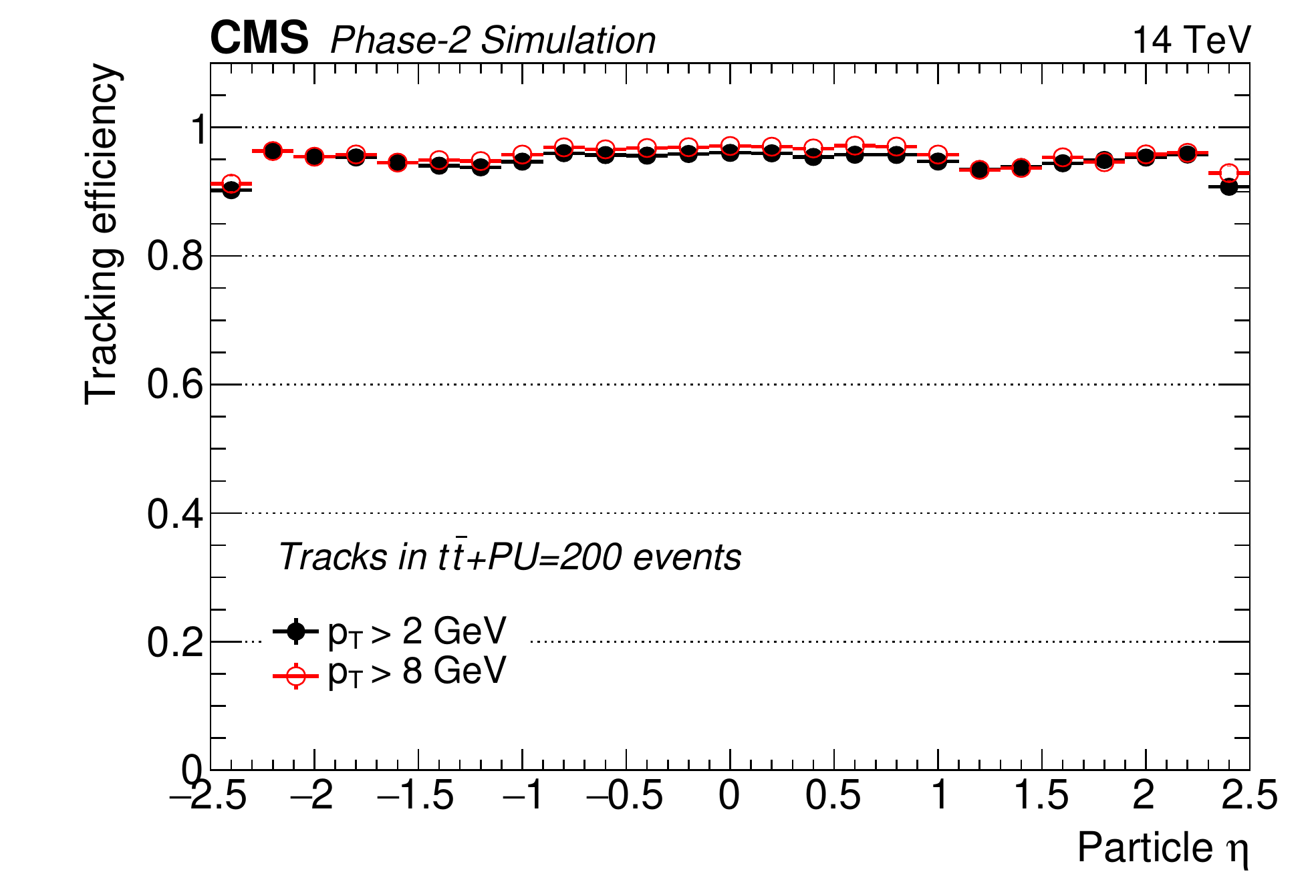}
\end{minipage}
\begin{minipage}[t]{0.49\textwidth}
\includegraphics[width=\textwidth]{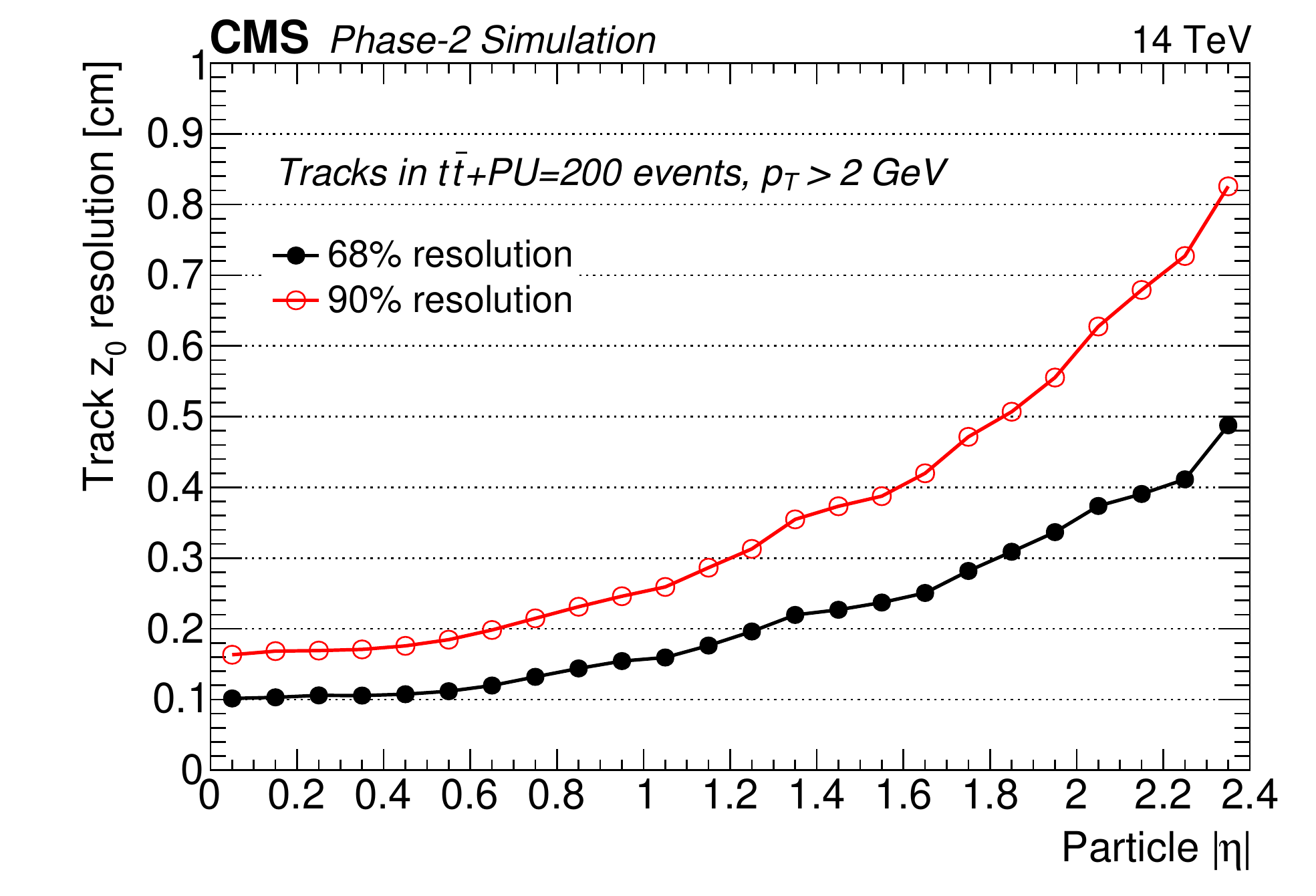}
\end{minipage}
\begin{minipage}[t]{0.4\textwidth}
\hspace{10mm}\includegraphics[width=\textwidth]{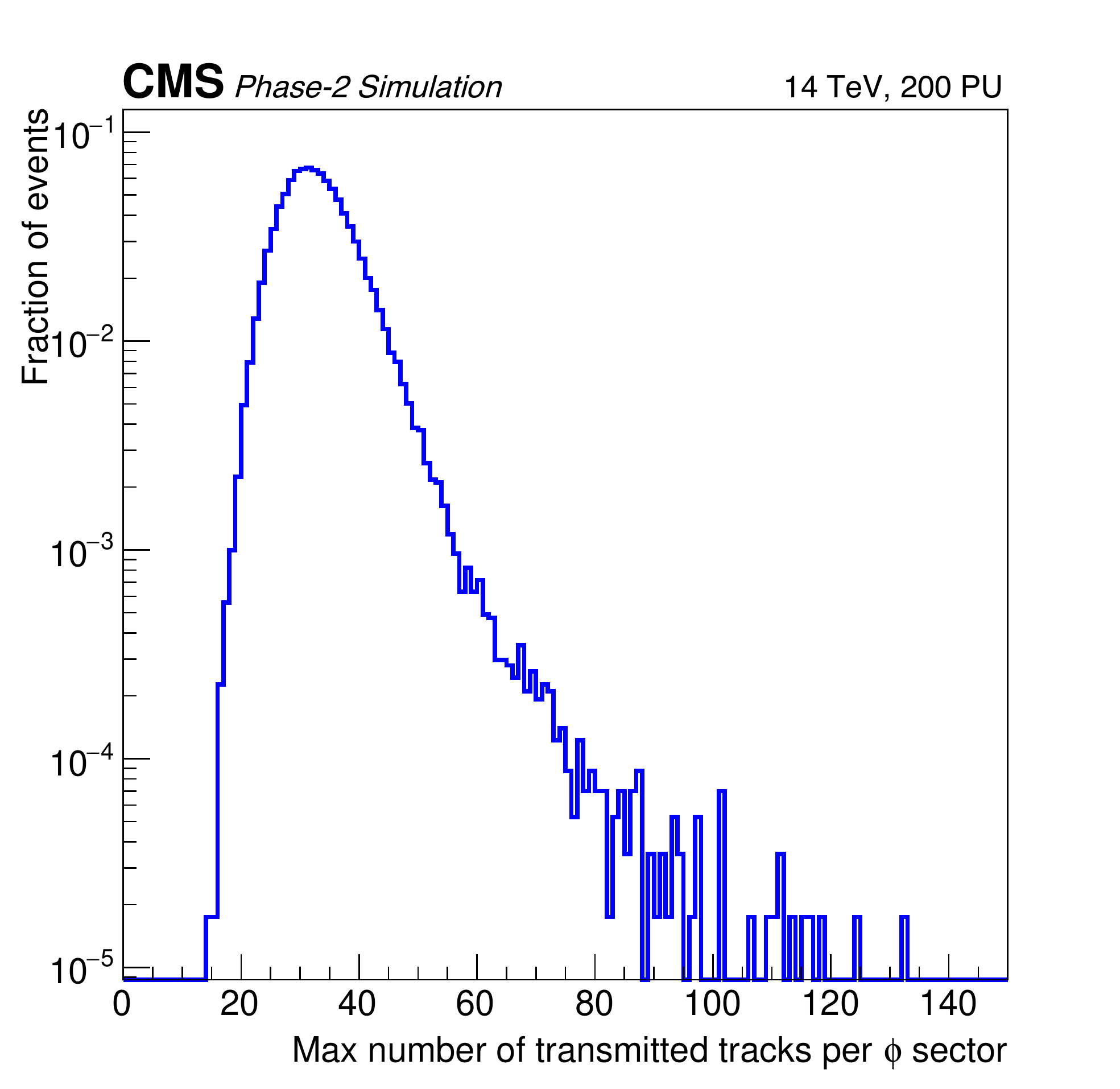}
\end{minipage}
\hspace{10mm}
\begin{minipage}[t]{0.49\textwidth}
\includegraphics[width=\textwidth]{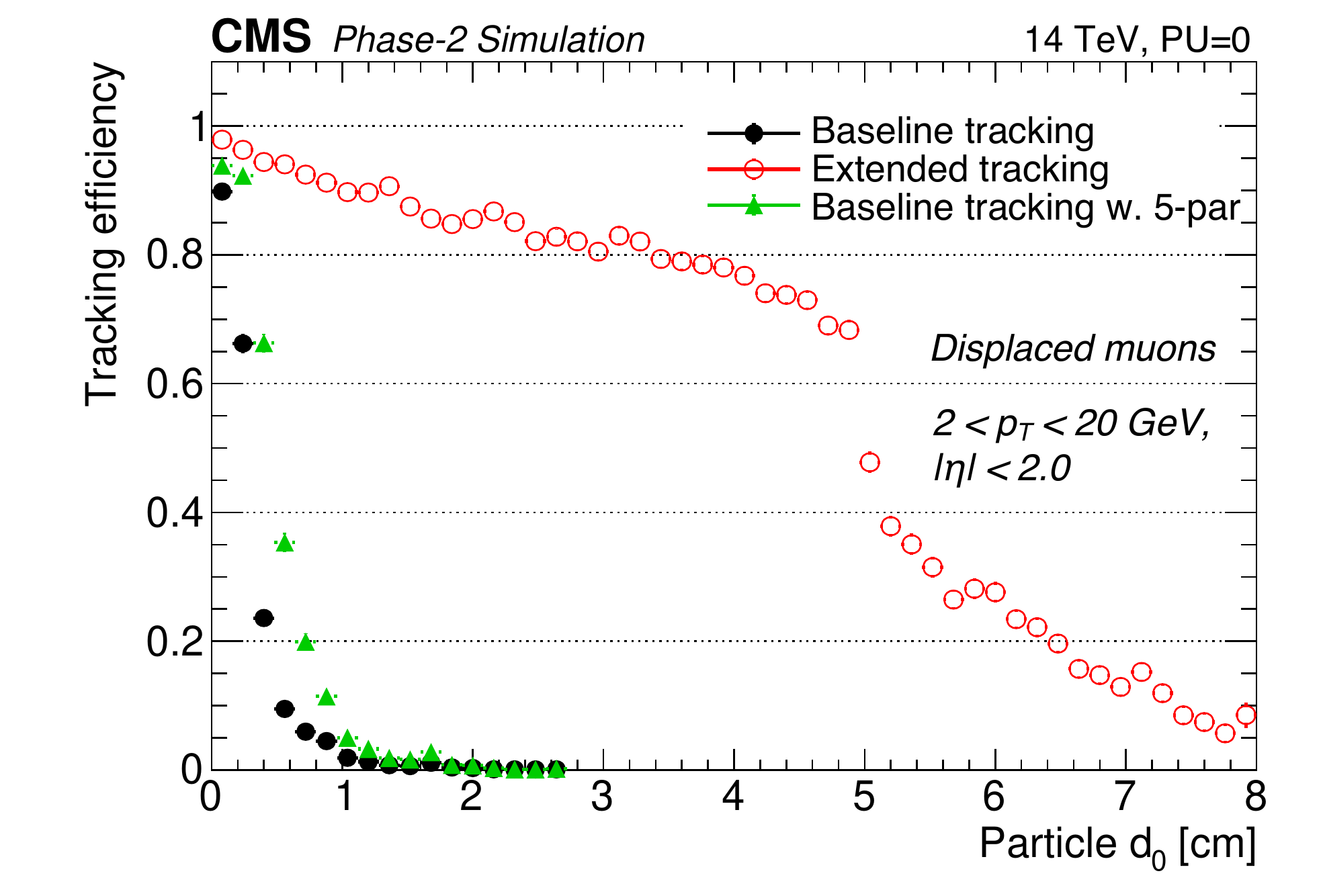}
\end{minipage}
\caption{L1 track finding efficiency as a function of particle $\eta$ (top left) and track longitudinal impact parameter resolution as a function of $|\eta|$ (top right) for charged particles from top quark pair events with an average of 200 additional pileup interactions; 
L1 track rate per $\phi$ sector shown for a minimum $\pt$ threshold of 2~GeV (bottom right); 
L1 track finding efficiency as a function of transverse impact parameter for tracks corresponding to long-lived particles (bottom left), showing the possible enhancement in efficiency from the extended, displaced tracking~\cite{CMS_P2_L1TDR}.}
\label{fig:L1TK_perf}
\end{center}
\end{figure}

%%%%%%%%%%%%%%%%%%%%%%%%%%%%%%%%%%%%%%%%%%%%%%%%%%%
\section{ATLAS HARDWARE-BASED TRACKING FOR HL-LHC}
\label{sec:outlook}

The ATLAS experiment has designed a new tracker for the HL-LHC 
upgrade~\cite{ATLAS_TDR_ITKStrip,ATLAS_ITK_Performance}. Similar to CMS, the original ATLAS tracker must be 
replaced as it cannot handle the
much higher data rates and radiation exposure at the HL-LHC. The layout of the
ATLAS tracker for the HL-LHC is shown in Fig.~\ref{fig:ATLAS_ITK}. It is based on an 
all-silicon design with pixel sensors in the inner regions where the occupancies
are highest and strip sensors in the outer regions. The acceptance of the strip detectors
covers the range $|\eta|<2.7$ while the pixels extend coverage to $|\eta|<4.0$.
Stereo angles are implemented in the strip detectors to provide a second coordinate measurement. 

\begin{figure}[tbp]
\begin{center}
\includegraphics[width=0.80\textwidth]{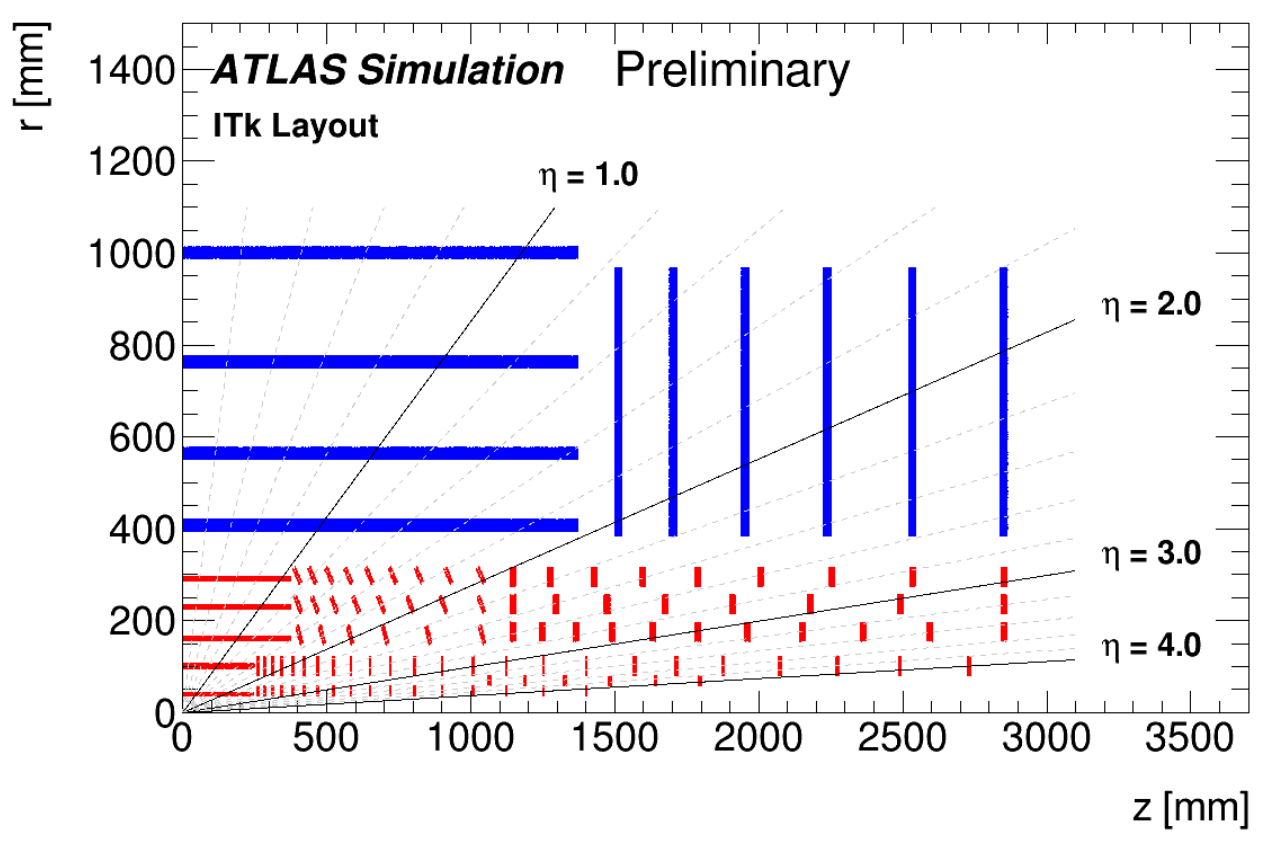}
\caption{\label{fig:ATLAS_ITK} 
The proposed geometry for the ATLAS inner tracker upgrade for the HL-LHC is an all Si based design. The innermost region, $r<350$~mm, is instrumented with pixel detectors in a configuration with five barrel layers and four rings in the forward region. The radial region $350<r<1000$~mm is instrumented with strip detectors with four barrel layers and six forward disks. From Ref.~\cite{ATLAS_ITK_Performance}. 
}
\end{center}
\end{figure}

The ATLAS experiment's use of tracking in the trigger system differs from that of CMS described earlier 
in that the ATLAS tracker does not provide trigger primitives. Hence ATLAS has chosen a different approach to hardware-based tracking where full-detector trajectory reconstruction is not performed in the first stage of the trigger system. The ultimate trigger system design is under development; here we present the assumptions outlined in Ref.~\cite{atlas_TDAQ_TDR}. 

An overview of the ATLAS trigger system is shown in Fig.~\ref{fig:ATLAS_TDAQ}, where both the baseline architecture with a single-level hardware trigger and the evolved architecture with a two-level hardware trigger, also referred to as an "L0/L1" architecture, are shown. The baseline architecture operates a single Level-0 (L0) hardware trigger using information from the calorimeter and muon detector systems, with a maximum readout rate of 1~MHz and latency of 10~$\mu$s. ATLAS is additionally maintaining upgrade projects with flexibility of possibly including L1 track reconstruction through the L0/L1 architecture, where regional tracking is operating at an input rate of 4~MHz. The evolved architecture is in particular intended to mitigate risks associated with uncertainties in hadronic trigger rates and occupancies in the inner pixel detector layers~\cite{atlas_TDAQ_TDR}. 

The primary goal of incorporating tracking in the ATLAS trigger is to reconstruct tracks in hardware as an input to the HLT to reduce CPU usage. It is foreseen to use pattern recognition with AMs (see Sect.~\ref{sec:AMPR}), in combination with track fitting in FPGAs, though different options are under consideration. In the baseline architecture, the so-called Hardware Tracking for the Trigger (HTT) is expected to perform a combination of regional (seeded) and global (full-detector) tracking. Regional tracking (rHTT) would be performed for approximately 10\% of the detector with an input event rate of 1~MHz, seeded by the L0 trigger that utilizes information from the calorimeter and muon systems only, while the global full-detector tracking (gHTT) would be performed as a second step, with an input rate of 100~kHz. The HTT system is largely based on the developments toward the FTK system that was intended to be used by ATLAS in LHC Run-3~\cite{FTK}. The rHTT is expected to reconstruct trajectories with $\pt > 2$~GeV, whereas the gHTT would target $\pt > 1$~GeV, both covering the acceptance region of $|\eta|<4.0$. Since the HTT is not operating in the first-level trigger, the latency constraints are less stringent than for the CMS L1 track trigger system, and the allowed latency is of the order of 100~$\mu$s. 

The HTT system is foreseen to be based on 48 independent HTT units, providing track reconstruction for a dedicated $\eta-\phi$ region, each corresponding to one ATCA shelf. The track finding is divided in two steps, implemented in common Tracking Processor (TP) boards. The first stage performs the pattern recognition through AM ASICs hosted in TP boards with Pattern Recognition Mezzanines, and in the second stage, the track fitting is carried out in FPGAs in TP boards with Track Fitting Mezzanines. 

An alternative FPGA-based, Hough Transform implementation of the pattern recognition is being studied as an alternative to the use of the Associative Memory ASIC. 
Both the baseline AM-based as well as the FPGA-based implementations provide flexibility to enhance the selection of long-lived particles~\cite{ATLAS_HTT_displaced}.

\begin{figure}[tbp]
\begin{center}
\includegraphics[width=0.99\textwidth]{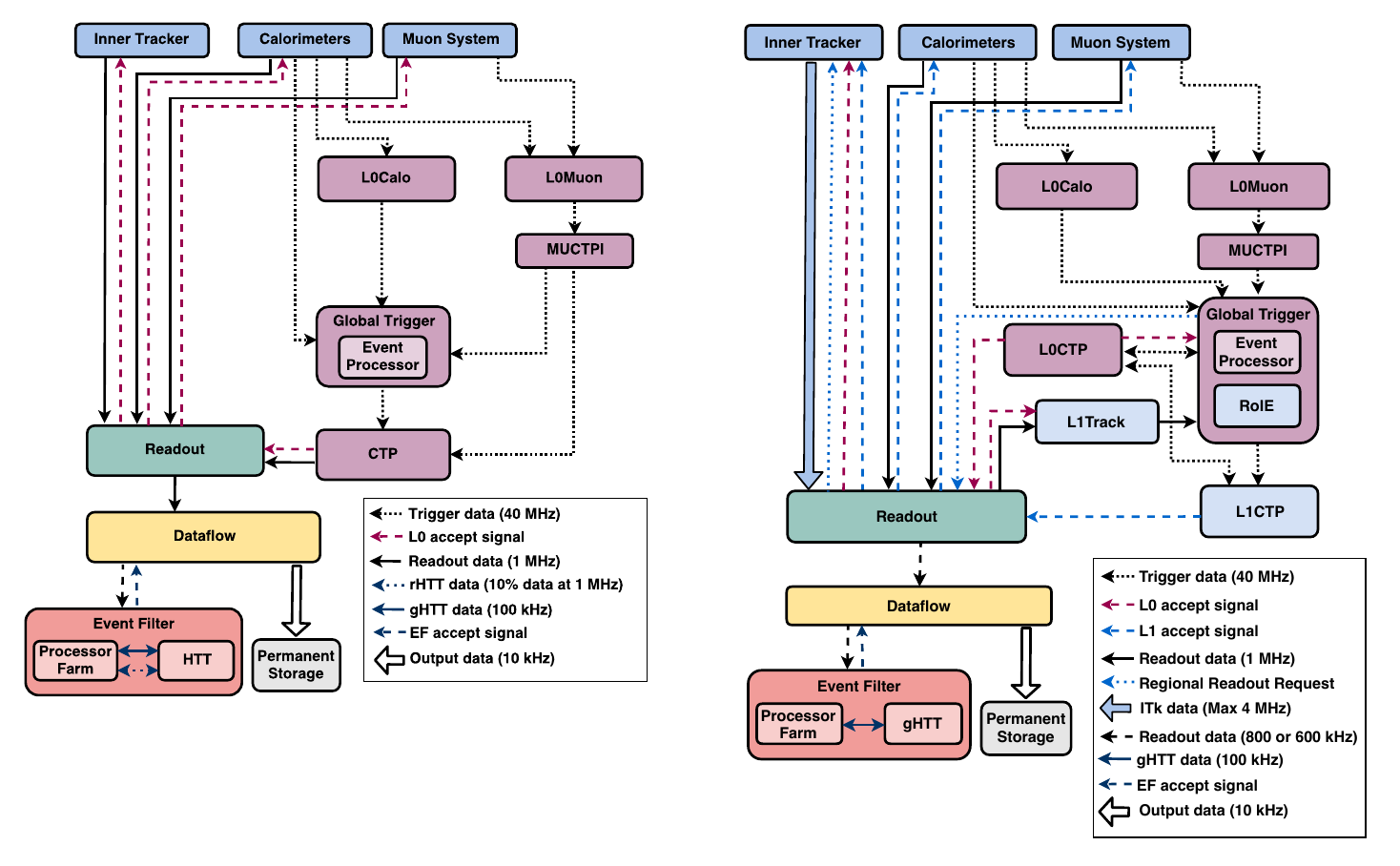}
\caption{\label{fig:ATLAS_TDAQ} 
Diagram of the ATLAS baseline TDAQ system (left) with a single-level hardware trigger and the evolved TDAQ system (right) with a two-level hardware trigger. From Ref.~\cite{atlas_TDAQ_TDR}. 
}
\end{center}
\end{figure}

%%%%%%%%%%%%%%%%%%%%%%%%%%%%%%%%%%%%%%%%%%%%%%%%%%%
\section{SUMMARY} 
\label{sec:summary}

One of the main challenges to fully explore the HL-LHC potential involves maintaining sufficiently low thresholds to efficiently trigger on the important electroweak physics processes
in the high pileup environment. Including information from the charged-particle tracking detectors in the hardware trigger
will provide a major new handle to control the trigger rates while
maintaining thresholds to efficiently trigger on $H/W/Z$ boson production. The inclusion of tracking 
improves the quality of almost all trigger objects, including
muon, electrons, hadronic $\tau$s, jets, missing transverse energy, and energy sums. 

The CMS Collaboration has designed a new tracker for the HL-LHC
upgrades where providing trigger primitives is a key 
novel capability. The upgraded outer tracker will be constructed using 
a new type of modules, $\pt$ modules, which consist of two
Si sensors spaced a few millimeters apart. By correlating hits in
the two sensors in logic implemented on the module, trigger 
primitives (stubs) can be formed. 
These stubs, required to be consistent with charged particles with a transverse momentum above 2 GeV, can be read out at the full bunch crossing rate of 40~MHz. They serve as the input to a backend L1 track finding system. 

CMS has demonstrated the ability to efficiently reconstruct 
L1 tracks in dedicated hardware within the latency requirements 
of the L1 trigger. Three different approaches were pursued for the pattern recognition plus track fitting: the tracklet road search algorithm, the Hough 
transform plus Kalman Filter, and the Associative Memory approach. 
All three approaches make use of time multiplexing, where 
data from different bunch crossings are distributed to different 
processing units, and a spatial division of the detector. 
Each of these approaches were 
shown to have the potential to work and CMS is 
pursuing an all-FPGA based approach that uses the tracklet 
road search algorithm for the pattern recognition, combined with the 
Kalman filter for the final fit. 

Prototypes for the hardware for the L1 tracking, the Serenity 
and Apollo boards, have been developed. The Serenity 
board will implement the DTC functionality, which will unpack and 
organize the data into the correct processing boards and time 
multiplexing slices. The Apollo board will implement the 
track finding algorithm.

The ATLAS Collaboration is pursuing a tracker design optimized
for offline reconstruction
for their HL-LHC upgrades. In the baseline trigger architecture,
hardware-based track reconstruction is used as an input to the
software-based high-level trigger. The implementation is planned to
use associative memories for pattern finding and FPGA-based linearized
track fitting. An alternative architecture is envisaged, where low-latency
regional track reconstruction is included as part of the L1 hardware trigger with an input rate of 4~MHz.

Hardware-based, low latency track reconstruction will provide an
important handle for the triggers at HL-LHC. The use of
track information will provide an essential tool to mitigate the
effects of high pileup. Significant work has been performed to
demonstrate the feasibility of these approaches for the planned
detector designs, as documented in the experiments' technical
design reports for the HL-LHC upgrades. These upgrade projects 
are now proceeding to the implementation stage. The hardware 
will be installed in the experiments during LHC upgrades in 
2025-2027 and operations with the upgraded detectors are
scheduled to start in late 2027.

%Disclosure
\section*{DISCLOSURE STATEMENT}
The authors were early proponents of the development of the CMS track
trigger and the tracklet approach in particular. 

% Acknowledgements
\section*{ACKNOWLEDGMENTS}
The authors thank their CMS collaborators and ATLAS friends for useful
discussions and insights. A.R.~is grateful to the U.S.~National
Science Foundation for its
continued support under grant NSF-PHY-1912813.

% References

\bibliographystyle{ar-style5}
\bibliography{TT_AR}

\end{document}